\documentclass[a4paper,11pt]{article}

\usepackage{jcappub}
\usepackage{amssymb}
\usepackage{graphicx}
\usepackage{amsmath}
\usepackage{subcaption}
\usepackage{multirow}
\usepackage{multicol}
\usepackage{makecell}
\usepackage{bm}
\usepackage{journal_abbre}

\newcommand{\md}{\text{d}}

\makeatletter

\newcommand{\Rmnum}[1]{\expandafter\@slowromancap\romannumeral #1@}
\makeatother

\def\Msun{{\rm M}_{\odot}}

\title{Cross-correlating Squared kSZ and HI Intensity Fields: A Map-level ACT--MeerKAT Forecast}

\author[1,2]{Yu-Er Jiang,}
\author[1,2]{Yun Liu,}
\author[1,2,3*]{Yan Gong,}
\emailAdd{(Corresponding authors) $^{\ast}$gongyan@bao.ac.cn}
\author[4]{Zi-Yan Yuwen,}
\author[5\dag]{Yin-Zhe Ma,}\emailAdd{$^{\dagger}$mayinzhe@sun.ac.za}
\author[1]{Furen Deng,}
\author[1,2,6,7,8]{Xuelei Chen,}
\author[9,10,11]{Qi Guo}

\affiliation[1]{National Astronomical Observatories, Chinese Academy of Sciences, 20A Datun Road, Beijing 100101, China}
\affiliation[2]{School of Astronomy and Space Science, University of Chinese Academy of Sciences (UCAS), Yuquan Road No.19A, Beijing 100049, China}
\affiliation[3]{Science Center for Chinese Space Station Survey Telescope, National Astronomical Observatories, Chinese Academy of Sciences, 20A Datun Road, Beijing 100101, China}
\affiliation[4]{Asia Pacific Center for Theoretical Physics (APCTP), Pohang 37673, Korea}
\affiliation[5]{Department of Physics, Stellenbosch University, Matieland 7602, South Africa}
\affiliation[6]{Department of Physics, College of Sciences, Northeastern University, Shenyang 100819, China}
\affiliation[7]{Centre of High Energy Physics, Peking University, Beijing 100871, China}
\affiliation[8]{State Key Laboratory of Radio Astronomy and Technology, China}
\affiliation[9]{Institute for Frontiers in Astronomy and Astrophysics, Beijing Normal University, Beijing 102206, China}
\affiliation[10]{School of Physics and Astronomy, Beijing Normal University, Beijing 100875, China}
\affiliation[11]{Key Laboratory for Computational Astrophysics, National Astronomical Observatories, Chinese Academy of Sciences, Beijing 100101, China}

\abstract{The kinetic Sunyaev--Zel'dovich (kSZ) effect probes the line-of-sight momentum of free electrons, but its blackbody spectrum and velocity-dependent sign make it difficult to isolate and cause conventional cross-correlations with density tracers to vanish. In this work, we investigate the cross-correlation of the squared kSZ field with the projected squared H\textsc{i} intensity field using mock ACT and MeerKAT observations constructed from a Jiutian-1G simulation lightcone. We generate halo-based kSZ maps and MeerKAT-like H\textsc{i} brightness-temperature cubes over a $13.5^\circ\times13.5^\circ$ field. The simulations incorporate ACT beam smoothing and effective internal-linear-combination noise, together with the frequency-dependent MeerKAT beam, scanning strategy, map-making, and inhomogeneous thermal noise. We analyze signal-only, idealized amplitude-reconstructed, and Wiener-filtered kSZ maps. {The resulting squared-field cross-correlations have diagnostic cumulative signal-to-noise ratios of approximately $3$--$4$.} Our map-level results demonstrate the detectability of the $\mathrm{kSZ}^2\times\mathrm{H\textsc{i}}^2$ correlation in future joint ACT--MeerKAT observations.
}

\begin{document}
\maketitle
\flushbottom

\section{Introduction}

Observations of the cosmic microwave background (CMB) constitute one of the three observational pillars of the standard $\Lambda$ cold dark matter ($\Lambda$CDM) cosmological model \citep{Efstathiou:1990xe}. Over the past two decades, increasingly sensitive ground-based CMB experiments, including the Atacama Cosmology Telescope (ACT) \citep{Choi2020,ACTPol:2015teu}, the Simons Observatory (SO) \citep{SimonsObservatory:2025wwn}, and the South Pole Telescope (SPT) \citep{Camphuis2025}, have achieved the high angular resolution and instrumental sensitivity required for precise measurements of secondary CMB anisotropies. One such secondary anisotropy is the kinetic Sunyaev--Zel'dovich (kSZ) effect, which is generated by Thomson scattering of CMB photons by free electrons with bulk peculiar motion \citep{ZeldovichSunyaev1969,SunyaevZeldovich1980}. The resulting temperature fluctuation encodes the line-of-sight integral of the electron density weighted by the peculiar velocity, making the kSZ effect a distinctive probe of the matter distribution and velocity field in the late-time Universe.

Cross-correlation with tracers of large-scale structure provides an effective means of extracting the information carried by the kSZ signal. Existing approaches include pairwise kSZ statistics based on spectroscopic tracers \citep{Hand2012,Planck2016-unbound,Ma2017,Li2018}, velocity-weighted reconstruction \citep{CHM2015,Planck2016-unbound,Hadzhiyska:2024qsl}, and correlations with projected density fields \citep{Hill2016,Ferraro2016}.
Particularly, neutral-hydrogen 21 cm intensity mapping provides a complementary, redshift-resolved view of large-scale structure. By measuring the aggregate line emission from unresolved galaxies, this technique efficiently maps the H\textsc{i} brightness-temperature field over large cosmological volumes. A range of radio facilities have conducted or planned 21 cm intensity-mapping surveys, including the Square Kilometre Array (SKA) \citep{Santos2015SKAHIIM,Yohana2019,Wang2021MeerKAT19,Cunnington2022MeerKAT19,Spinelli2022MeerKLASS}, Parkes \citep{Anderson2018Parkes,Tramonte2019,Tramonte2020}, the Green Bank Telescope (GBT) \citep{Chang2010,Masui2013,Wolz2022GBT}, the Canadian Hydrogen Intensity Mapping Experiment (CHIME) \citep{Newburgh2014CHIMECalibration,Bandura2014CHIMEPathfinder}, the Five-hundred-meter Aperture Spherical radio Telescope (FAST) \citep{Bigot2016,SmootDebono2017FAST,Yohana2019,Yohana2021,Hu2021,Deng2022FASTCSST}, Tianlai \citep{Chen2011Tianlai,Perdereau2022Tianlai}, and the Baryon Acoustic Oscillations from Integrated Neutral Gas Observations (BINGO) experiment \citep{Dickinson2014BINGO,Bigot2015,Battye2016,Yohana2019,Zhang2022BINGO,Abdalla2022,Wuensche2022}.

Among these facilities, MeerKAT has already demonstrated the potential of single-dish intensity mapping and cross-correlation analyses at low redshift \citep{Wang2021MeerKAT19,Cunnington2022MeerKAT19,Meerklass2025}. At the same time, the arcminute angular resolution of ACT makes its CMB maps well suited to kSZ studies \citep{Choi2020,Naess2025}. Combining these data sets is physically compelling since ionized electrons and neutral hydrogen trace distinct baryonic phases embedded in the same underlying matter and velocity fields.


In this work, we develop a map-level simulation of the cross-correlation between the squared kSZ field and the line-of-sight projection of the squared H\textsc{i} field. We investigate this $\mathrm{kSZ}^2\times\mathrm{H\textsc{i}}^2$ statistic using a cosmological lightcone constructed from the Jiutian-1G simulation \citep{Han2025Jiutian}. Free electrons are assigned to halos using an NFW-based model \citep{Luchina2025DEMNUni}, while the H\textsc{i} masses are obtained from a semi-analytic galaxy model \citep{Obreschkow2009}. We then construct a $13.5^\circ\times13.5^\circ$ mock field with ACT-like and MeerKAT-like angular responses. On the CMB side, we consider three inputs: the intrinsic beam-smoothed kSZ map; an idealized amplitude-reconstructed branch consisting of the kSZ signal plus effective internal-linear-combination (ILC) noise; and a Wiener-filtered blackbody branch containing the lensed primary CMB, kSZ signal, and ILC noise. On the H\textsc{i} side, we simulate a MeerKAT L-band intensity-mapping observation that includes a frequency-dependent beam, the survey scanning pattern, map-making, and inhomogeneous thermal noise. Constructing both observables from the same lightcone preserves their physical correlation while allowing the principal map-level effects to be isolated.

Besides, we pay particular attention to the order in which squaring and angular downgrading are applied. Because the ACT-like kSZ map is sampled much more finely than the H\textsc{i} map, we compare downgrading the kSZ map before squaring with squaring the high-resolution map before downgrading. These operations retain different amounts of unresolved signal and noise variance and are therefore not mathematically interchangeable. Finally, we compare the auto- and cross-spectra of the resulting squared fields and report diagnostic cumulative signal-to-noise ratios for the simulated patch.

The paper is organized as follows: Section~\ref{sec:simulation} describes the Jiutian lightcone and the electron and H\textsc{i} models; Section~\ref{sec:mock_map} presents the ACT-like CMB and MeerKAT-like intensity-map simulations;  Section~\ref{sec:hi_projected_fields} defines the squared and projected fields, including the two kSZ resolution-ordering choices; Section~\ref{sec:spectra} presents the flat-sky power spectra and diagnostic signal-to-noise estimates; the summary and relevant discussions are given in Section~\ref{sec:summary_discussion}.

\section{Simulation and Model}
\label{sec:simulation}

\subsection{Jiutian-1G simulation and lightcone}
\label{subsec:jiutian}

Because the kSZ effect is a projected signal accumulated along CMB photon trajectories, we use a lightcone simulation to generate mock maps of both the kSZ effect and H\textsc{i} intensity mapping. The lightcone is assembled from Jiutian-1G snapshots spanning {$0.0015\lesssim z\lesssim 4$}. Jiutian-1G is the $1\,h^{-1}{\rm Gpc}$ simulation box within the Jiutian suite.
Jiutian is a state-of-the-art hybrid cosmological simulation suite {\citep{Han2025Jiutian}} run with the \textsc{LGadget-3} code and developed to support extragalactic science with the Chinese Space-station Survey Telescope (CSST) 
\citep{Zhan2011CSST,Cao2018CSSOS,Gong2019CSSOS,Zhan2021CSST,CSSTCollaboration2026}. Dark matter halos are identified using the friends-of-friends (FoF) algorithm \citep{Davis1985}, and gravitationally bound substructures are identified with \textsc{SubFind} \citep{Davis1985,Springel2001}. Halo merger trees are subsequently constructed with the B-Tree code. Jiutian-1G adopts a $\Lambda$CDM cosmology with parameters consistent with the Planck 2018 results. The simulation and cosmological parameters are listed in Table~\ref{tab:jiutian}. The \textsc{LGalaxies} code \citep{Henriques2015} is then used to implement a semi-analytic model (SAM) of galaxy formation and predict the galaxy properties required for mock survey catalogs.

\begin{table}
    \begin{center}
    \caption{Parameters of the Jiutian lightcone simulation.}
    \label{tab:jiutian}
    \begin{tabular}{lc}
        \hline\noalign{\smallskip}
        Parameter & Value \\
        \hline\noalign{\smallskip}
        Box side length & $1000\,h^{-1}{\rm Mpc}$ \\
        Number of particles & $6144^3$ \\
        Particle mass & $3.723\times10^8\,h^{-1}\Msun$ \\
        Force softening & $4.0\,h^{-1}{\rm kpc}$ \\ 
        $\Omega_{\rm m}$ & 0.3111 \\
        $\Omega_{\Lambda}$ & 0.6889 \\
        $\Omega_{\rm b}$ & 0.0490 \\
        $n_{\rm s}$ & 0.9665 \\
        $\sigma_8$ & 0.8102 \\
        $h$ & 0.6766 \\
    \noalign{\smallskip}\hline
    \end{tabular}
    \begin{minipage}{0.92\textwidth}
    \end{minipage}
    \end{center}
\end{table}

\begin{figure}[t] 
    \centering
    \includegraphics[width=0.92\textwidth]{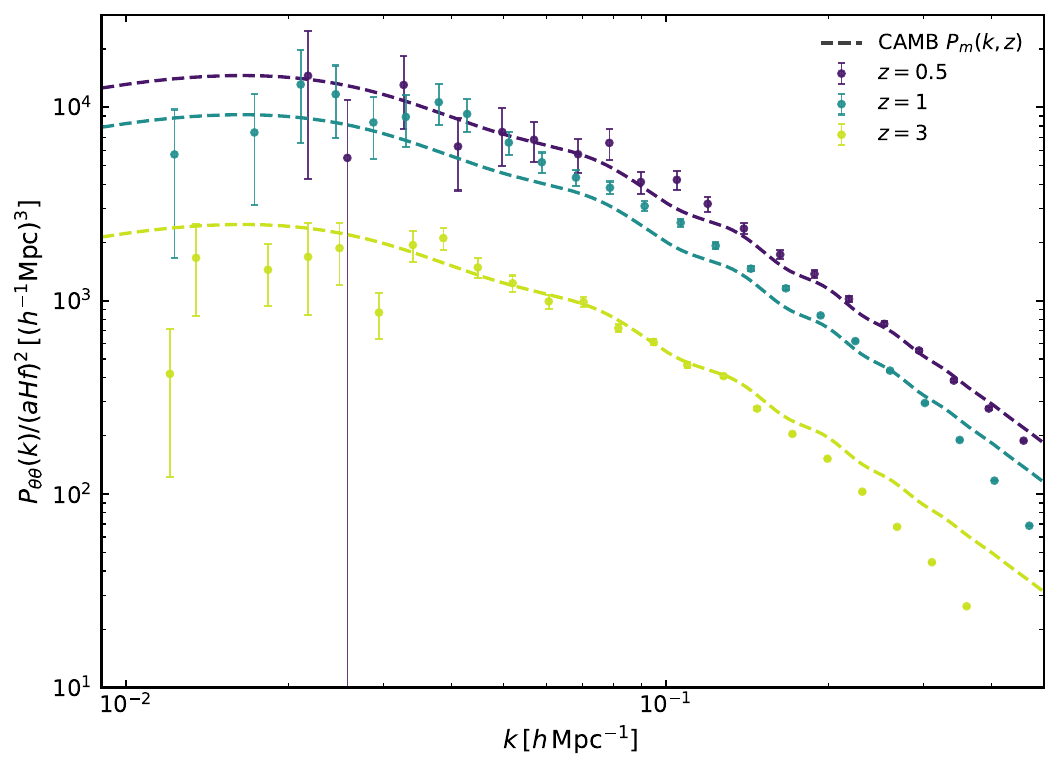}
    \caption{Velocity-divergence power spectra measured from the Jiutian lightcone at representative redshifts, compared with linear-theory matter power spectra computed using \textsc{CAMB}. We plot the velocity-divergence power spectrum after dividing by $(aHf)^2$, for which linear theory predicts 
    $P_{\theta\theta}/(aHf)^2\simeq P_{\rm m}$.
    The comparison is used as a lightcone-level sanity check for the velocity field that enters the kSZ map construction.}
    \label{fig:lightcone_velocity_divergence}
\end{figure}

Unlike the H\textsc{i} intensity field alone, the kSZ signal depends explicitly on the accuracy of the velocity field reproduced by the cosmological simulation. In the linear regime, the power spectrum of the velocity divergence, $\theta\equiv-\nabla\cdot{\bm v}$, is proportional to the matter power spectrum:
\begin{equation}
    P_{\theta\theta}(k,z)
    = \left[a(z) H(z) f(z)\right]^2
    P_{\rm m}(k,z)~,
    \label{eq:velocity_divergence_linear_theory}
\end{equation}
where $a$ is the scale factor, $H$ is the Hubble parameter, and $f=\md\ln D/\md\ln a$ is the linear growth rate.
Figure~\ref{fig:lightcone_velocity_divergence} compares $P_{\theta\theta}/(aHf)^2$ measured from the Jiutian lightcone with the matter power spectrum $P_{\rm m}$ computed using \textsc{CAMB} \citep{Lewis2000CAMB}.
{On small scales, the measured lightcone velocity-divergence spectra are suppressed relative to the CAMB theory curves. We interpret this suppression mainly as a numerical damping effect associated with the finite grid resolution in the power spectrum estimation. 
The effective comoving pixel scale increases with redshift, so the damping becomes more visible at higher redshift. 
We didn't deconvolve the corresponding window function to recover the amplitude, in order to avoid introducing additional numerical errors. }
This comparison provides a consistency check of the velocity field that enters the kSZ simulation. 

\subsection{Electron Distribution}
\label{subsec:electron_model}

As a secondary CMB anisotropy generated by Thomson scattering of CMB photons by free electrons along the line of sight \citep{Kompaneets1957,ZeldovichSunyaev1969,SunyaevZeldovich1970,Sunyaev1977}, the kSZ effect requires an explicit model of the electron distribution. We adopt a halo-based electron model \citep{Luchina2025DEMNUni}, in which the free-electron number density is derived from the ionized gas density within each halo:
\begin{equation}
    n_{\rm e}(r) = 
    \frac{\rho_{\rm gas}(r)}{\mu m_{\rm p}},
    \label{eq:electron_density_profile}
\end{equation}
where $m_{\rm p}$ is the proton mass and {$\mu = \left[f_{\mathrm{H}} + \frac{1}{2}(1-f_{\mathrm{H}})\right]^{-1}$} is the mean molecular weight per free electron in fully ionized gas, where $f_{\rm H}\simeq 0.75$ is the hydrogen mass fraction. Eq.~(\ref{eq:electron_density_profile}) assumed that both hydrogen and helium are fully ionized in the gas component of baryons. The baryonic gas density follows an NFW profile \citep{Navarro1997},
\begin{equation}
    \rho_{\rm gas}(r)
    =
    (1-f_*) f_{\mathrm{b}}\,
    \frac{\rho_0}{(r/r_{\rm s})(1+r/r_{\rm s})^2},
    \label{eq:ksz_baryon_nfw}
\end{equation}
where $f_*=0.1$ is the stellar mass fraction and $f_{\mathrm{b}}=\Omega_\mathrm{b}/\Omega_{\mathrm{m}}$ is the halo baryon fraction. The scale density $\rho_0$ is obtained by normalizing $\rho(r)$ to $M_{200}$ within $R_{200}$. The dependence of the scale radius $r_{\rm s}$ on $M_{200}$ and redshift is incorporated through the concentration--mass relation of Duffy et al. (2008)~\cite{Duffy2008}:
\begin{equation}
    c_{200}
    \equiv
    \frac{R_{200}}{r_{\rm s}}
    =
    5.71
    \left(
    \frac{M_{200}}{2\times10^{12}\,h^{-1}\Msun}
    \right)^{-0.084}
    (1+z)^{-0.47}.
    \label{eq:duffy_concentration}
\end{equation}

\subsection{HI model}

\label{subsec:hi_model}
In our simulation, each mock galaxy in the lightcone catalog is assigned a cold-gas mass, cold-gas metallicity, and stellar-disk mass by \textsc{L-Galaxies}. These properties are used to determine the H\textsc{i} mass following the semi-analytic prescription of \citep{Obreschkow2009}. The hydrogen in the cold gas is assumed to be either neutral atomic hydrogen, H\textsc{i}, or molecular hydrogen, ${\rm H}_2$, with most of the cold gas residing in axisymmetric, thin galactic disks. For a galaxy with cold-gas mass $M_{\rm cg}$ and cold-gas metal mass $M_Z$, the H\textsc{i} mass is 
\begin{equation}
    M_{{\rm{H \textsc{i}}}}
    =
    (M_{\rm cg}-M_Z)\,
    \beta\,
    (1+R_{\rm H_2})^{-1},
    \label{eq:mhi_partition}
\end{equation}
where $\beta=0.75$ is the hydrogen mass fraction and $R_{\rm H_2}\equiv M_{\rm H_2}/M_{H\textsc{i}}$ is the molecular-to-atomic hydrogen mass ratio. Assuming the exponential gas profiles inferred from observations \citep{BlitzRosolowsky2006,Leroy2008}, the integrated ratio $R_{\rm H_2}$ is given by
\begin{equation}
    R_{\rm H_2}
    \simeq
    \left[
    3.44\left(R_{\rm H_2}^{\rm c}\right)^{-0.506}
    +
    4.82\left(R_{\rm H_2}^{\rm c}\right)^{-1.054}
    \right]^{-1},
    \label{eq:rh2_integrated}
\end{equation}
where the central value of the radial molecular-to-atomic hydrogen ratio is
\begin{equation}
    R_{\rm H_2}^{\rm c}
    =
    \left[
    \frac{
    G\,M_{\rm cg}\,
    }{
    8\pi P_{\rm mid} r_{\rm disk}^4
    }\left(M_{\rm cg}+\langle f_\sigma\rangle M_{\star,{\rm disk}}\right)
    \right]^{0.8}.
    \label{eq:rh2_central}
\end{equation}
Here, $G$ is the gravitational constant, $M_{\star,{\rm disk}}$ and $r_{\rm disk}$ are the stellar-disk mass and gas-disk scale length from the Jiutian simulation, respectively, $P_{\rm mid}=2.35\times10^{-13}$ is an empirical parameter describing the kinematic mid-plane pressure \citep{Leroy2008}, and $\langle f_\sigma\rangle$ is the assumed ratio of the vertical velocity dispersions of gas and stars \citep{Dickey1990,Bottema1993,Leroy2008}.

\section{Mock Map Making}
\label{sec:mock_map}

Having obtained the free-electron and neutral-hydrogen distributions in the Jiutian lightcone, we convert both fields into mock observable maps. The lightcone has a circular cross-section with a radius of $10^\circ$. To facilitate power-spectrum estimation, we extract a square $13.5^\circ\times13.5^\circ$ region, corresponding to a simulated survey area of approximately $182.25\,\mathrm{deg}^2$. We then apply ACT- and MeerKAT-like instrumental responses to the kSZ and H\textsc{i} maps, respectively, in order to approximate the relevant observational effects.

ACT is a ground-based millimeter-wave CMB experiment located in the Atacama Desert and designed to measure temperature and polarization anisotropies at arcminute resolution. Its recent data products provide multifrequency CMB maps over a large fraction of the sky. The DR6 maps are based on Advanced ACTPol observations in bands centered near 98, 150, and 220 GHz \citep{Choi2020,Naess2025}. The latest DR6.02 internal-linear-combination (ILC) products include Compton-$y$, CMB blackbody-temperature, and CMB blackbody E-mode polarization maps, sampled at $0.5'$ resolution with a $1.6'$ Gaussian beam \citep{Naess2025}.

MeerKAT is a 64-dish SKA-precursor radio telescope in South Africa and is well suited to single-dish H\textsc{i} intensity mapping. The MeerKLASS program was designed to map the aggregate redshifted 21 cm emission over wide areas with MeerKAT, enabling large-scale-structure and cosmological analyses based on H\textsc{i} intensity maps \citep{Santos2017}. Here, we simulate MeerKAT L-band intensity mapping over $900$--$1015\,\mathrm{MHz}$, corresponding to $0.40\lesssim z\lesssim0.58$. 
{This frequency interval is chosen to center our analysis at \(z\simeq0.5\), close to the redshift range where MeerKAT L-band intensity maps have already enabled significant cross-correlation measurements with optical galaxy surveys \citep{Cunnington2022MeerKAT19,Meerklass2025}.}

\subsection{ACT-like CMB maps}
\label{subsec:act_like_cmb}

The fractional CMB temperature perturbation from the kSZ effect is
\begin{equation}
    \frac{\Delta T_{\rm kSZ}(\hat{\bm n})}{T_{\rm CMB}}
    =
    -\sigma_{\rm T}
    \int d\chi\,
    n_{\rm e}(\chi\hat{\bm n})\,
    e^{-\tau}
    \frac{v_{\parallel}(\chi\hat{\bm n})}{c},
    \label{eq:ksz}
\end{equation}
where $\sigma_{\rm T}$ is the Thomson cross-section, $n_{\rm e}$ is the free-electron number density, $v_{\parallel}$ is the line-of-sight peculiar velocity, $c$ is the speed of light, $\chi$ is the comoving distance, and $\tau$ is the optical depth which is given by
\begin{equation}
    \tau = 
    \sigma_{\rm T}
    \int^{z}_{0}\,
    \frac{\bar{n}_{\rm e}(z')}{1+z'}\,
    \frac{c~\mathrm{d}z'}{H(z')},
    \label{eq:tau}
\end{equation}
where $\bar{n}_{\rm e}(z)$ is the mean electron density at redshift $z$. Because the lightcone covers only the optically thin redshift range, $z\leq6$, we adopt $e^{-\tau}\simeq1$ when calculating Equation~\ref{eq:ksz}.

\begin{figure}[t]
    \centering
    \begin{subfigure}{0.48\textwidth}
        \centering
        \includegraphics[width=\textwidth]{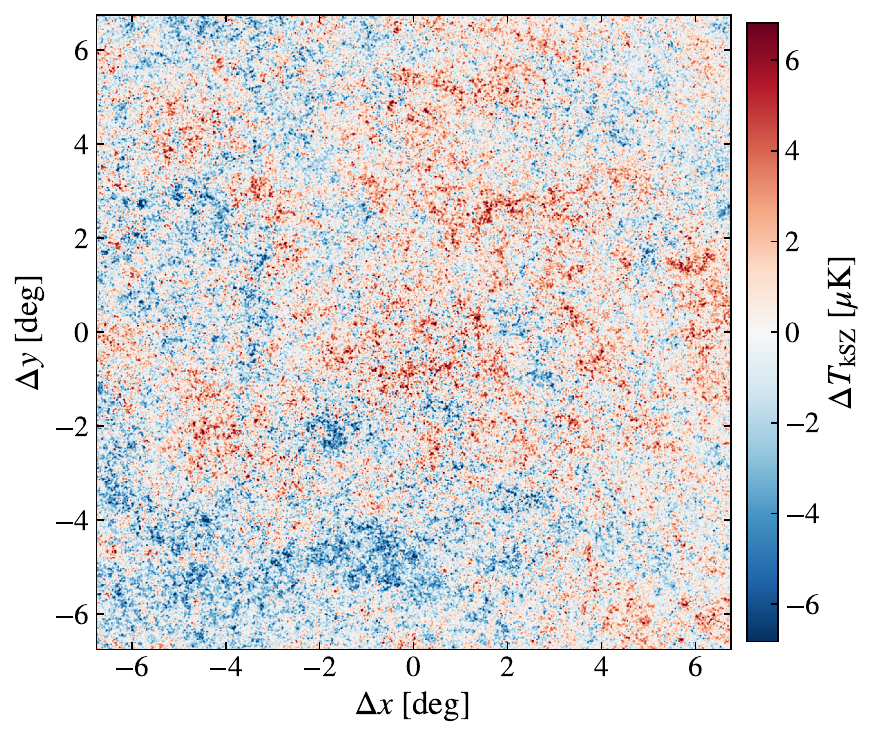}
        \label{fig:observed_ksz_map}
    \end{subfigure}
    \hfill
    \begin{subfigure}{0.48\textwidth}
        \centering
        \includegraphics[width=\textwidth]{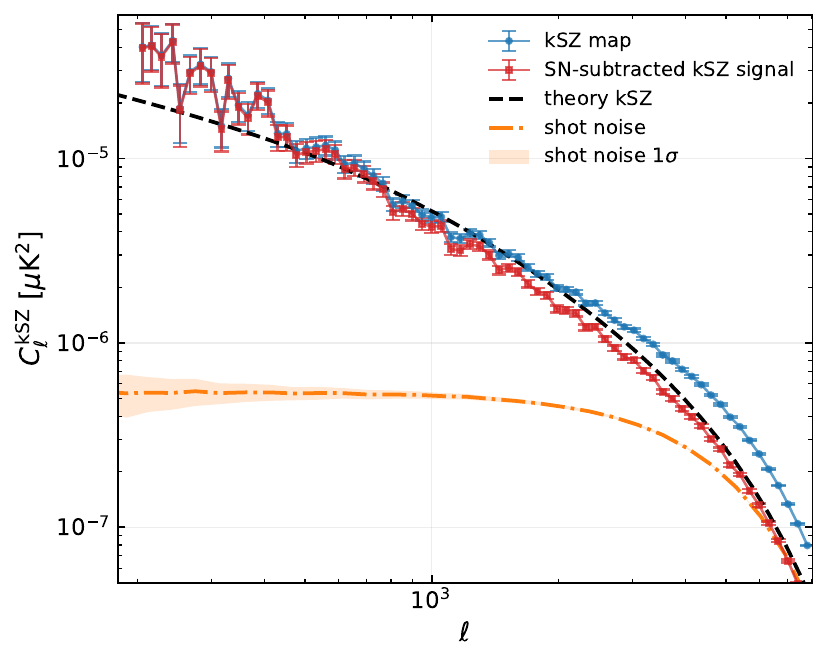}
        \label{fig:ksz_auto_shot_noise}
    \end{subfigure}
    \caption{$\it Left\ panel$: the kSZ temperature map smoothed with a $1.6'$ beam. $\it Right\ panel$: the auto-power spectrum of the simulated map before and after subtraction of the shot-noise contribution. The blue and red curves show the raw and shot-noise-subtracted simulated spectra, respectively, the black dashed curve is the theoretical kSZ reference spectrum, and the orange curve and shaded region show the mean and scatter of the shot-noise estimates obtained from randomized-position realizations. The $1\sigma$ region is the sample standard deviation of 100 such realizations.}
    \label{fig:ksz_map_and_auto}
\end{figure}

We set the angular pixel size of the kSZ map to $0.5'$ to match the latest ACT ILC products \citep{Naess2025}. For each pixel, $\Delta T_{\mathrm{kSZ}}/T_{\mathrm{CMB}}$ is evaluated by integrating the contributions of electrons in all intervening halos along the line of sight according to Equation~\ref{eq:ksz}. 
We multiply this fractional temperature distortion by the CMB blackbody temperature $T_{\rm CMB}$ and convolve the resulting kSZ temperature map with a Gaussian beam of FWHM $1.6'$ to match the ACT DR6.02 NILC temperature products \citep{Naess2025}. The resulting kSZ temperature map and the estimated kSZ auto-power spectrum are shown in the left and right panels of Figure~\ref{fig:ksz_map_and_auto}, respectively.

Because our kSZ signal is generated from discrete, halo-associated electron components rather than from a continuous hydrodynamical electron field, the auto-spectrum contains a shot-noise-like contribution arising from the finite halo sampling. 
We estimate this contribution using random-position catalogs that preserve the halo kSZ weights while redistributing their angular positions within the same footprint. Each randomized catalog is processed through the same map-making, beam-smoothing, and power-spectrum pipeline as the signal catalog. The right panel of Figure~\ref{fig:ksz_map_and_auto} compares the high-resolution kSZ auto-spectrum with both the theoretical kSZ spectrum and the randomized-position shot-noise estimate. The simulated spectrum lies above the theoretical curve on small angular scales, where the shot-noise contribution is non-negligible. As expected for a Poisson-like white-noise component, the estimated shot noise is approximately scale-independent in $C_\ell$; its suppression on the smallest angular scales is caused by the ACT Gaussian beam of $\theta_{\rm FWHM} = 1.6'$ by a factor of $\left(B_{\ell}^{\rm A}\right)^2$, which can be expressed as
\begin{equation}
    B_\ell^{\rm A}
    =
    \exp\left[-\frac{\ell(\ell+1)\sigma_{\rm A}^2}{2}\right],
    \qquad
    \sigma_{\rm A}
    =
    \frac{\theta_{\rm FWHM}}{\sqrt{8\ln 2}}.
    \label{eq:act_gaussian_beam}
\end{equation}
After subtracting the contribution of shot noise, the simulated kSZ auto-spectrum agrees well with the theoretical spectrum computed over the same redshift range.

Unlike the thermal SZ effect, the kSZ signal has the same blackbody frequency dependence as the primary CMB and therefore cannot be isolated through conventional frequency-based ILC methods. A common approach is to apply a Wiener filter in harmonic space to suppress primary CMB anisotropies, other secondary contributions, and instrumental noise relative to the kSZ signal \citep{Yuwen2025,Hill2016,Ferraro2016,Dore2004}. The Wiener filter is defined as
\begin{equation}
    F_\ell=\frac{C_\ell^{\mathrm{kSZ,sig}}}{C_\ell^{\mathrm{tot}}}.
    \label{eq:Wiener filter}
\end{equation}
where $C_\ell^{\mathrm{kSZ,sig}}$ and $C_\ell^{\mathrm{tot}}$ are the angular power spectra of the kSZ signal and total CMB temperature field, respectively. At map level, however, Wiener filtering also suppresses the kSZ amplitude by a factor {$C_\ell^{\mathrm{kSZ,sig}}/C_\ell^{\mathrm{tot}}$}. We therefore construct two mock kSZ observables. The first is an amplitude-reconstructed kSZ map, based on the assumption that a future component-separation method can effectively remove foregrounds while recovering the kSZ amplitude. The second is a Wiener-filtered kSZ map. To construct these maps, we additionally simulate the relevant CMB components and effective instrumental noise.

\begin{figure}[t] 
    \centering
    \begin{subfigure}{0.31\textwidth}
        \centering
        \includegraphics[width=\textwidth]{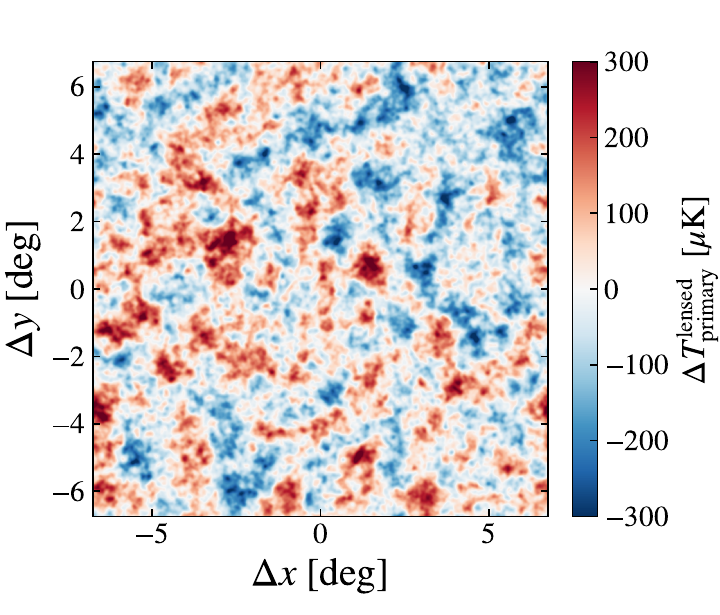}
    \end{subfigure}
    \hspace{0.04\textwidth}
    \begin{subfigure}{0.31\textwidth}
        \centering
        \includegraphics[width=\textwidth]{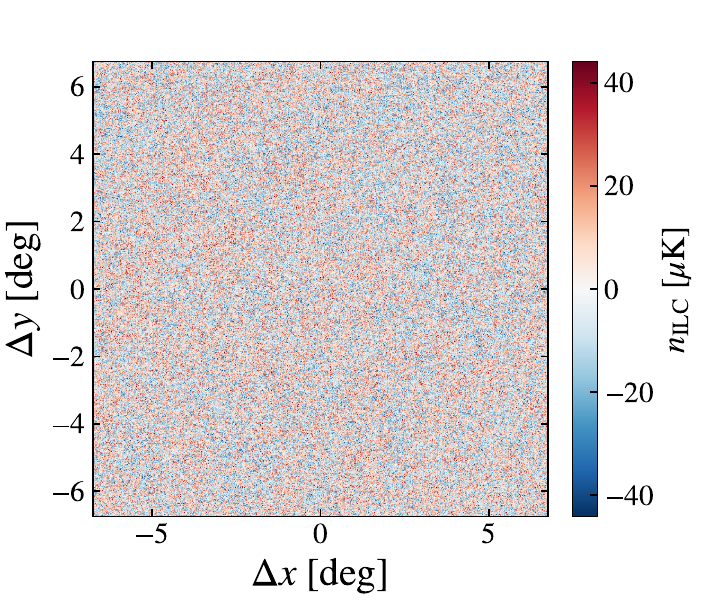}
    \end{subfigure}
    \vspace{0.6em}
    
    \begin{subfigure}{0.31\textwidth}
        \centering
        \includegraphics[width=\textwidth]{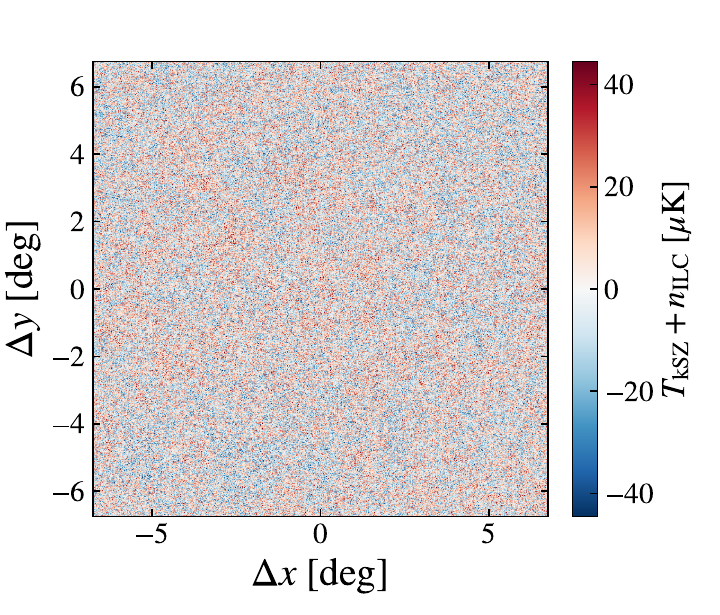}
    \end{subfigure}
    \hfill
    \begin{subfigure}{0.31\textwidth}
        \centering
        \includegraphics[width=\textwidth]{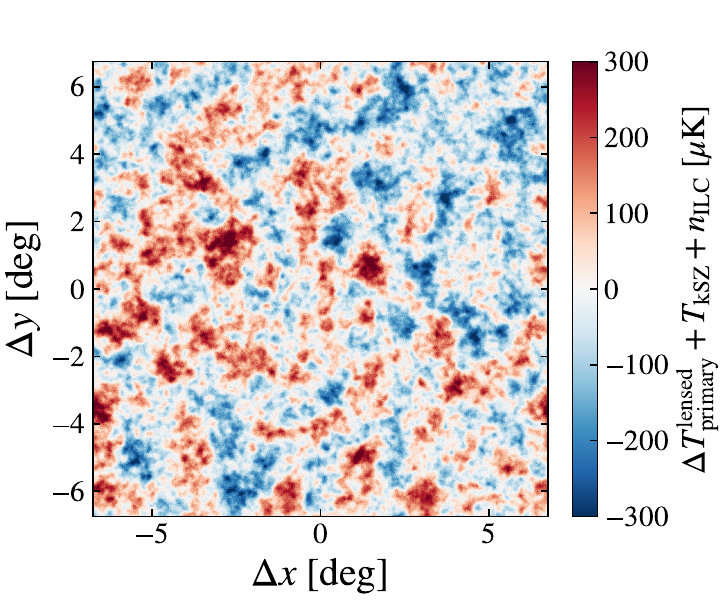} 
    \end{subfigure}
    \hfill
    \begin{subfigure}{0.31\textwidth}
        \centering
        \includegraphics[width=\textwidth]{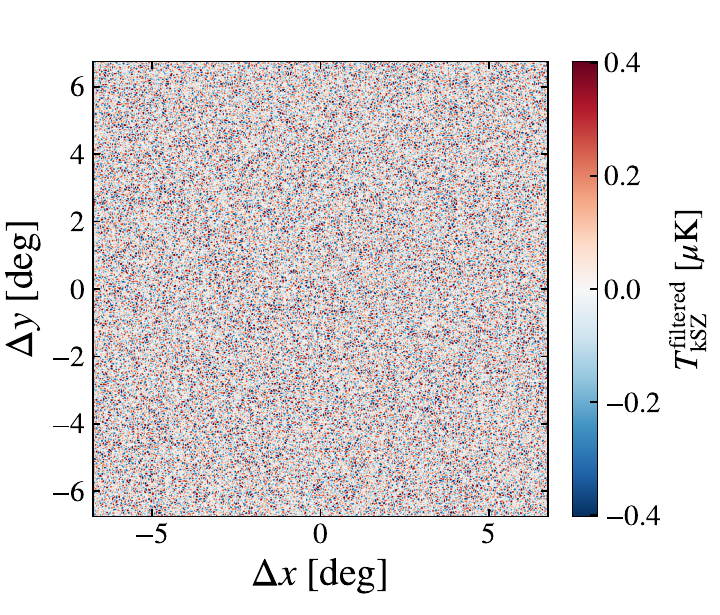}
    \end{subfigure}
    \caption{Components and derived maps of the mock ACT-like CMB observation on the $13.5^\circ\times13.5^\circ$ patch. The upper-left and upper-right panels show the lensed primary CMB and effective ILC-noise realizations, respectively. The lower-left panel shows the idealized amplitude-reconstructed branch, $T_{\rm kSZ}+n_{\rm ILC}$; the lower-middle panel shows the idealized ILC map, $T_{\rm ILC}$; and the lower-right panel shows the Wiener-filtered kSZ map constructed from the ILC CMB map.}
    \label{fig:act_like_cmb_components} 
\end{figure}

We define the amplitude-reconstructed kSZ map as
\begin{equation}
    T_{\rm kSZ}^{\rm re}(\hat{\bm n}) = T_{\rm kSZ}(\hat{\bm n}) + n_{\rm ILC}(\hat{\bm n}),
    \label{eq:ksz+noise}
\end{equation}
where $n_{\rm ILC}$ denotes the effective noise in the component-separated CMB observation. We generate a Gaussian realization from its angular power spectrum, $N_\ell^{\mathrm{ILC}}$, using the Python package \textsc{PowerBox} \citep{Murray2018Powerbox}. In a simple approximation, the noise power may be modeled as $N_\ell\simeq\Delta_T^2 (B^{\rm A}_{\ell})^{-2}$, where {$\Delta_T$} is the rms noise per radian and $B^{\rm A}_{\ell}$ is the ACT beam response in harmonic space. Instead, we adopt a more realistic model that captures the scale dependence of the post-ILC noise obtained from a weighted linear combination of multifrequency maps \citep{Yuwen2025}. We use the public package \textsc{orphics}\footnote{\url{https://github.com/msyriac/orphics}} to perform the harmonic-space ILC calculation and obtain the ACT noise spectrum $N^{\mathrm{ILC}}_\ell$ used in our analytical forecast \citep{Yuwen2025}. The instrumental parameters entering this calculation are listed in Table~\ref{tab:cmb_instrument_parameters}; see also the ACT DR6 data release \citep{Naess2025}. The resulting ILC-noise realization is shown in the upper-right panel of Figure~\ref{fig:act_like_cmb_components}.

\begin{table}[t] 
\begin{center}
\caption{ACT instrumental parameters used to evaluate the ILC noise power.}
\label{tab:cmb_instrument_parameters}
\begin{tabular}{cccc}
\hline\noalign{\smallskip}
Channel & Beam & $\delta T_N$ & Depth \\
\multicolumn{1}{c}{[GHz]} & \multicolumn{1}{c}{[arcmin]} & \multicolumn{1}{c}{[$\mu{\rm K}$-arcmin]} & \multicolumn{1}{c}{[mJy]} \\
\hline\noalign{\smallskip}
90  & 1.42 & 14  & 6.5 \\
150 & 2.07 & 14  & 8.4 \\
220 & 1.01 & 64  & 29 \\
\noalign{\smallskip}\hline
\end{tabular}
\end{center}
\end{table}

To construct the Wiener-filtered kSZ map, we first build an ACT-like ILC map. Public ACT DR6 component-separated maps use needlet ILC (NILC) methods to produce blackbody-temperature and Compton-$y$ products, including constrained variants in which selected foreground components are deprojected \citep{Coulton2024ACTNILC,Naess2025}. We therefore construct an idealized ACT ILC map under the assumption that all frequency-dependent components can be deprojected, leaving only components that preserve the CMB blackbody spectrum:
\begin{equation}
    T_{\rm ILC}(\hat{\bm n})
    =
    T_{\rm primary}^{\rm lensed}(\hat{\bm n})
    +
    T_{\rm ISW}(\hat{\bm n})
    +
    T_{\rm kSZ}(\hat{\bm n})
    +
    n_{\rm ILC}(\hat{\bm n}),
    \label{eq:t_obs}
\end{equation}
where $T_{\rm primary}^{\rm lensed}$ is the lensed primary CMB fluctuation and $T_{\rm ISW}$ denotes the integrated Sachs--Wolfe (ISW) contribution. The ISW signal is appreciable primarily at $\ell\lesssim100$, a range that cannot be measured adequately within our survey area. 
We therefore neglect the ISW contribution and use \textsc{PowerBox} to generate a lensed primary CMB realization from the lensed CMB power spectrum $C_\ell^{\rm lensed~CMB}$ computed with \textsc{CAMB}. 
The components of the mock ILC map are shown in Figure~\ref{fig:act_like_cmb_components}. The primary CMB and effective ILC noise dominate the map-level temperature fluctuations, while the kSZ signal has a substantially lower amplitude and is visually subdominant in both the amplitude-reconstructed and ILC maps.

\begin{figure}[t]
    \centering
    \begin{subfigure}{0.32\textwidth}
        \centering
        \includegraphics[width=\textwidth]{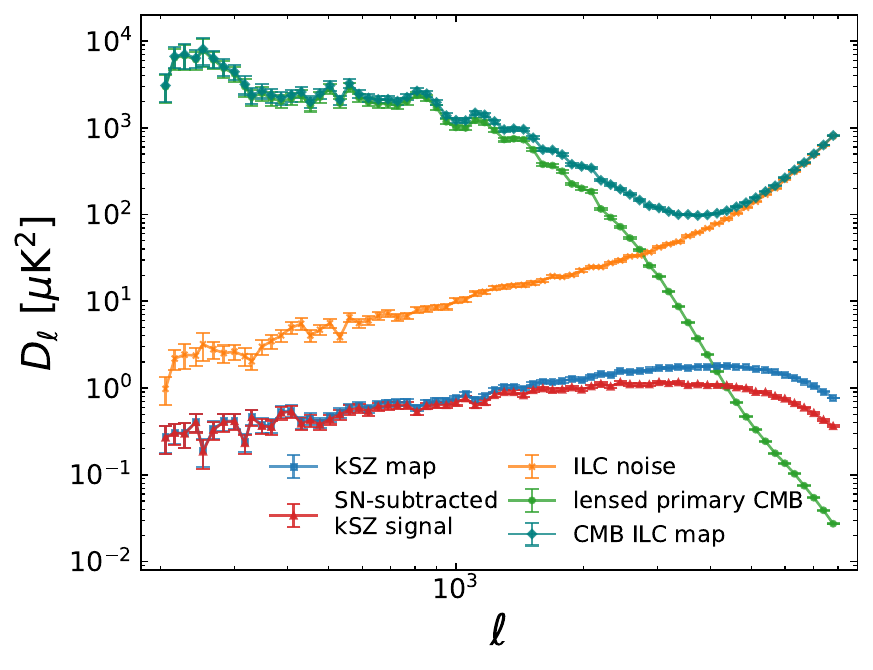}
    \end{subfigure}
    \hfill
    \begin{subfigure}{0.32\textwidth}
        \centering
        \includegraphics[width=\textwidth]{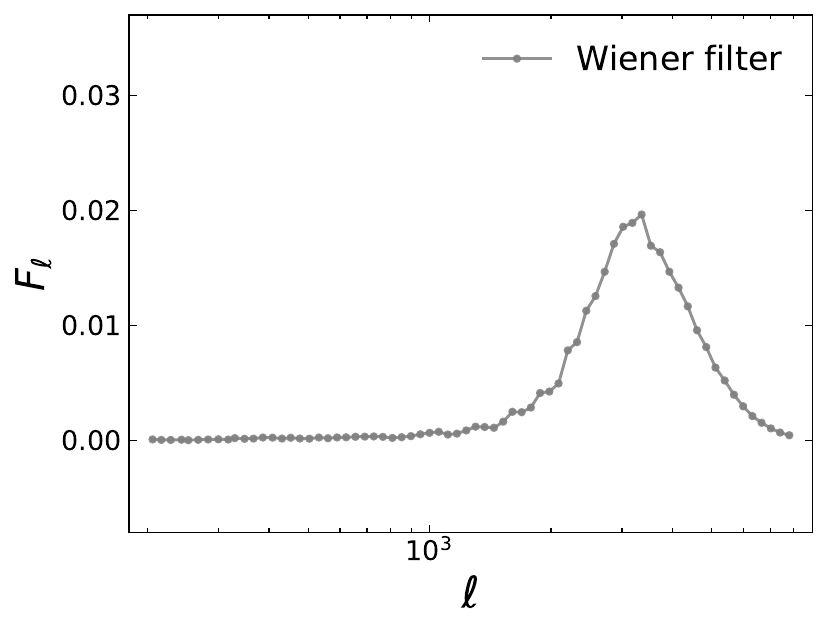}
    \end{subfigure}
    \hfill
    \begin{subfigure}{0.32\textwidth}
        \centering
        \includegraphics[width=\textwidth]{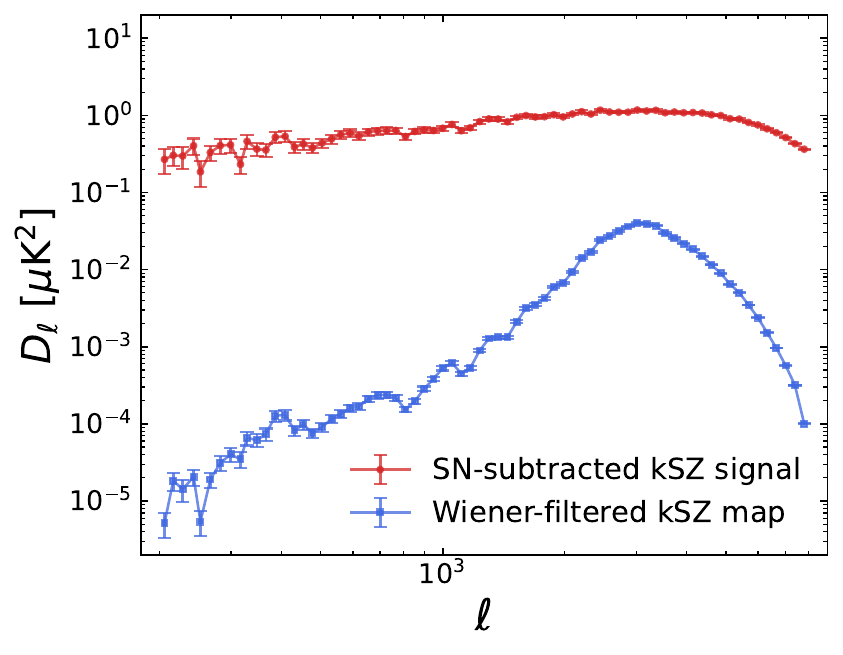}
    \end{subfigure}
    \caption{
    The left panel shows the auto-power spectrum of the simulated ACT ILC map, together with the auto-power spectra of its components including lensed primary CMB, effective ILC noise, and simulated kSZ map, as well as the shot-noise-subtracted kSZ signal adopted in the numerator of the Wiener filter. The middle panel shows the corresponding scale-dependent Wiener filter $F_\ell$ given by Equation~\ref{eq:Wiener filter used}. The right panel compares the auto-spectrum of the {shot-noise subtracted} kSZ signal with that of the Wiener-filtered kSZ map.
    } 
    \label{fig:wiener_filter_diagnostics}
\end{figure}

The Wiener filter adopted in this analysis is therefore
\begin{equation}
    F_\ell=\frac{C_\ell^{\mathrm{kSZ,sig}}}{C_\ell^{\mathrm{tot}}}
    =\frac{
    C_\ell^{\mathrm{kSZ}} - N_{\ell}^{\mathrm{SN}}
    }{
    C_\ell^{\rm lensed~CMB}
    +
    C_\ell^{\mathrm{kSZ}}
    +
    N^{\mathrm{ILC}}_\ell
    }.
    \label{eq:Wiener filter used}
\end{equation}
Here, $N_{\ell}^{\mathrm{SN}}$ is the kSZ shot-noise contribution in the simulation. This form serves the usual purpose of Wiener or matched filtering: it downweights modes dominated by the primary CMB and instrumental noise while retaining scales on which the relative kSZ contribution is larger \citep{Ferraro2016,Yuwen2025}. Figure~\ref{fig:wiener_filter_diagnostics} shows the resulting Wiener filter and the auto-power spectra $D_\ell\equiv\ell(\ell+1)C_\ell/(2\pi)$ of the components used to construct it. Consistent with the maps in Figure~\ref{fig:act_like_cmb_components}, the primary CMB and ILC noise dominate the total blackbody-temperature variance over most of the multipole range.
We apply the Wiener filter to the mock ILC map in harmonic space. Under the flat-sky approximation, the spherical-harmonic transform is replaced by a two-dimensional Fourier transform over the small sky patch. We Fourier-transform the mock ILC map, multiply each Fourier mode by the Wiener filter, and then apply the inverse transform to obtain the Wiener-filtered kSZ map:
\begin{equation}
    T_{\rm kSZ}^{\rm filtered}(\hat n)=\mathcal{F}^{-1}\{F_{\ell} \cdot \mathcal{F}\{T_{\rm ILC}(\hat n)\}\},
    \label{eq:filtered_map}
\end{equation}
where $\mathcal{F}$ and $\mathcal{F}^{-1}$ denote the forward and inverse Fourier transforms, respectively. The Wiener-filtered kSZ map is shown in the lower-right panel of Figure~\ref{fig:act_like_cmb_components}, and the right panel of Figure~\ref{fig:wiener_filter_diagnostics} compares its auto-spectrum with that of the original simulated kSZ signal. {The comparison illustrates the scale selection of the Wiener filter: it reduces the weight of modes dominated by primary CMB or ILC noise, while modes with larger relative kSZ contributions are preferentially retained, thereby suppressing contaminant-dominated modes and reducing the total variance in subsequent squared-field cross-correlation.}

\subsection{MeerKAT L-band H\textsc{i} intensity mapping}
\label{subsec:meerkat_like_hi}

\begin{figure}[t]
    \centering
    \begin{subfigure}{0.48\textwidth}
        \centering
        \includegraphics[width=\textwidth]{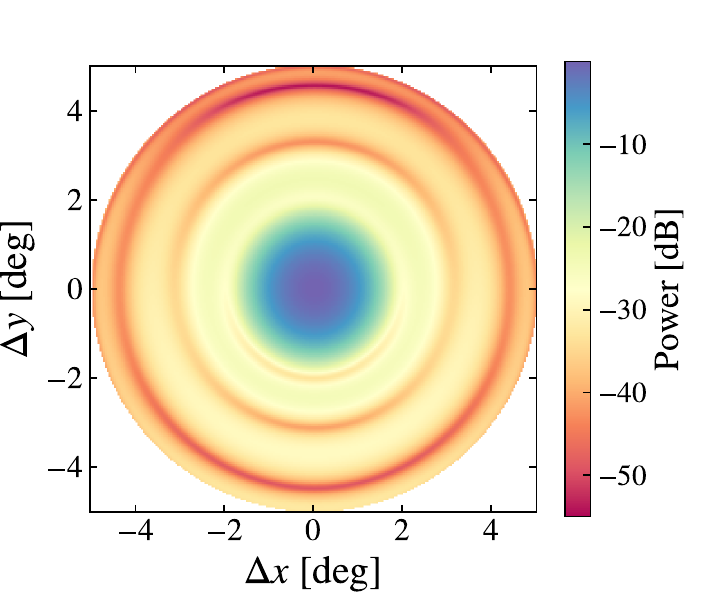}
    \end{subfigure}
    \hfill
    \begin{subfigure}{0.48\textwidth}
        \centering
        \includegraphics[width=\textwidth]{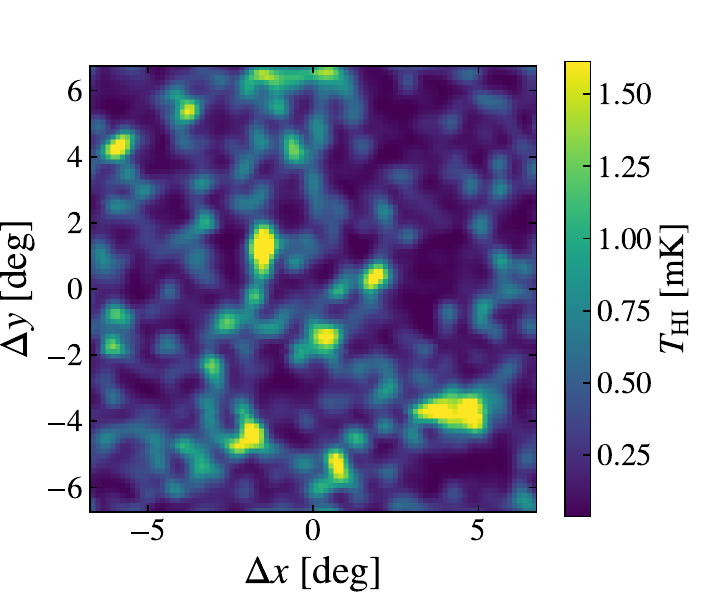}
    \end{subfigure}
    \vspace{0.5em}
    \begin{subfigure}{0.48\textwidth}
        \centering
        \includegraphics[width=\textwidth]{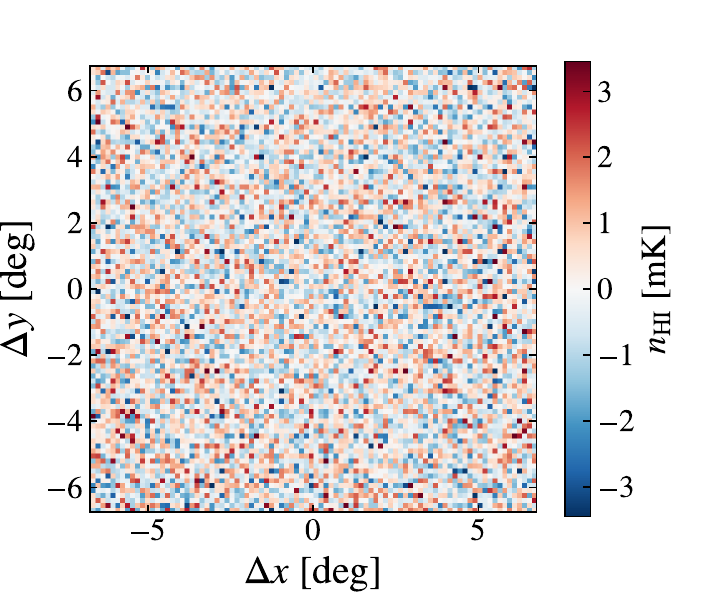}
    \end{subfigure}
    \hfill
    \begin{subfigure}{0.48\textwidth}
        \centering
        \includegraphics[width=\textwidth]{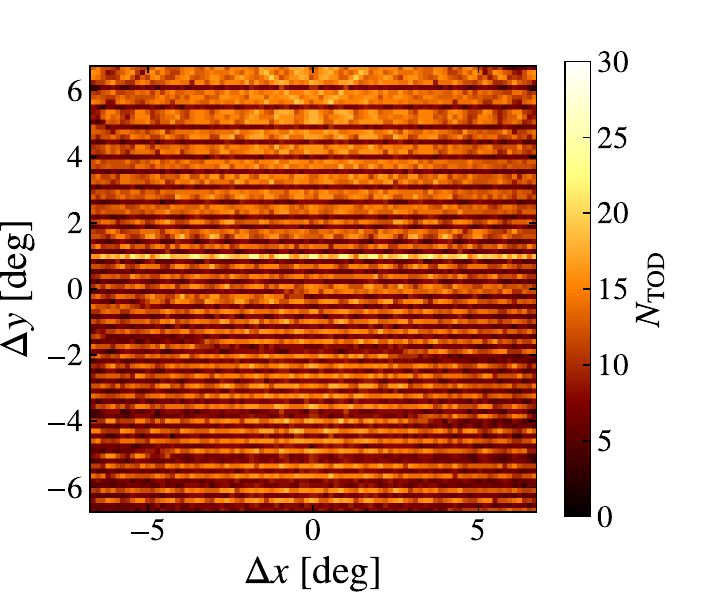}
    \end{subfigure}
    \caption{Components of the simulated MeerKAT L-band observation in the central frequency channel of $\nu\simeq957\,\mathrm{MHz}$. The upper-left panel shows the MeerKAT beam pattern generated with \textsc{EIDOS}, the upper-right panel shows the beam-smoothed H\textsc{i} intensity map, the lower-left panel shows the corresponding thermal-noise realization, and the lower-right panel shows the TOD number-count map, illustrating the inhomogeneous noise level produced by the on-the-fly scanning strategy.}
    \label{fig:hi_signal_noise_slice}
\end{figure}

We begin the MeerKAT H\textsc{i} intensity-map simulation by generating time-ordered data (TOD) for a specified survey strategy. We adopt the on-the-fly observing mode \citep{Rozgonyi2022} used by the completed MeerKAT L-band pilot survey (hereafter MeerKAT19) \citep{Wang2021MeerKAT19,Cunnington2022MeerKAT19} and the MeerKLASS L-band deep-field survey (hereafter MeerKAT21) \citep{Meerklass2025}. In this mode, the antennas scan in azimuth at a fixed elevation, reducing the effects of ground pickup and airmass variations on signal calibration. We use azimuth angles of $43^\circ$ and $62^\circ$ and a total observing time of 25 nights, ensuring that the scan tracks cover the full survey region. Following the MeerKAT19 observing strategy, each scan lasts 1.56 hours, with two symmetric scans performed per night. All 64 MeerKAT dishes follow the same scanning pattern, and the TOD are sampled every 2 seconds. The brightness temperature of each TOD sample is modeled as
\begin{equation}
    T_{\rm D}(\bm{\hat{n}},\nu)
    =
    \int d\bm r\,
    \mathcal{B}_{\rm M}(\bm{\hat{n}}-\bm r,\nu)\,
    T_{\rm b}(\bm r,\nu)
    +
    n_{\rm T},
    \label{eq:meerkat_tod}
\end{equation}
where $\mathcal{B}_{\rm M}(\bm{\hat{n}}-\bm r,\nu)$ is the frequency-dependent MeerKAT beam response, $T_{\rm b}(\bm r,\nu)$ is the sky brightness temperature at position $\bm r$, and $n_{\rm T}$ is the thermal-noise contribution. We generate the beam pattern in each frequency channel using the Python package \textsc{EIDOS}\footnote{\url{https://github.com/ratt-ru/eidos}} \citep{Asad2021MeerKATBeam}. Based on antenna holography measurements, \textsc{EIDOS} uses Zernike polynomials to model the MeerKAT L-band primary beam over a field with a maximum diameter of $10^\circ$. The model reproduces both the main lobe and the side lobes. The simulated primary-beam pattern is shown in the upper-left panel of Figure~\ref{fig:hi_signal_noise_slice}.

In general, $T_{\rm b}(\bm r,\nu)$ contains the H\textsc{i} signal $T_{\rm H\textsc{i}}$, Galactic and bright extragalactic radio foregrounds, and systematic contamination from radio-frequency interference (RFI), instrumental polarization leakage, and ground spillover \citep{Jiang2026MeerKATCSST}. In this work, for simplicity, we adopt the idealized case in which $T_{\rm b}$ contains only the H\textsc{i} signal. Radio foregrounds and instrumental contaminants in H\textsc{i} intensity mapping are typically orders of magnitude brighter than the cosmological signal and cannot be removed sufficiently through cross-correlation alone. Additional foreground-cleaning procedures inevitably cause some signal loss. For cross-correlations of unsquared fields, this loss can be calibrated using transfer functions constructed from ensembles of simulations and stochastic realizations, as demonstrated in previous analyses \citep{Anderson2018Parkes,Cunnington2022MeerKAT19,Meerklass2025}. This procedure relies on the signal loss being concentrated primarily in foreground-dominated modes and on the approximate independence of power-spectrum modes. Squaring the field couples Fourier modes through convolution, allowing signal loss that was initially confined to a restricted set of modes to propagate into otherwise signal-dominated modes. This mode coupling makes transfer-function calibration substantially more difficult. We therefore assume that the foregrounds are perfectly deprojected, while emphasizing that foreground removal and the associated signal-loss calibration remain essential challenges for applying the squared-field estimator to real observations. We defer a detailed treatment of these effects to future work.

As described in Section~\ref{subsec:hi_model}, the H\textsc{i} mass of each galaxy $M_{\rm H\textsc{i}}$ is obtained from the semi-analytic model of \citep{Obreschkow2009}. Then the H\textsc{i} distribution in the Jiutian lightcone $\rho_{\rm H\textsc{i}}$ can be derived from the position and H\textsc{i} mass of galaxies.
The H\textsc{i} brightness temperature at angular position $\bm{\hat n}$ and redshift $z$ is then given by \citep{Hall2013}
\begin{eqnarray}
    T_{\rm b}(\bm{\hat{n}},z) &=& \overline{T}_{\rm b}(z)\left(1+\Delta_{T_{\rm b}}(\bm{\hat{n}},z) \right),
\end{eqnarray}
where 
\begin{eqnarray}
    \overline{T}_{\rm b}(z)=188\,{\rm mK}\,h\Omega_{\rm HI}(z)\frac{(1+z)^{2}}{E(z)},
\end{eqnarray}
is the mean 21-cm brightness temperature evolution and $\Omega_{\rm HI}(z)$ is the comoving mass density of HI in units of the current critical density ($\rho_{\rm cr}$). $E(z)=H(z)/H_0$, where $H(z)$ is the Hubble parameter at redshift $z$ and $H_0$ is the Hubble constant. $\Delta_{T_{\rm b}}(\bm{\hat{n}},z)$ is the brightness temperature fluctuation, which depends on the HI number density and the peculiar velocity (also known as the redshift-space distortion)~\cite{Hall2013}. The redshift-space distortion is also considered by assigning galaxies to frequency channels according to their line-of-sight peculiar velocities. The mean H\textsc{i} brightness temperature of the resulting data cube in redshift range $0.40\lesssim z\lesssim0.58$ is found to be $0.35\,\mathrm{mK}$.

We model the instrumental thermal noise of the single-dish survey as Gaussian noise in each TOD. Its root-mean-square (rms) temperature is \citep{Bull2015IntensityMapping}
\begin{align}
    \sigma_{\rm T}=\frac{T_{\rm sys}}{\sqrt{\delta\nu t_{\rm scan}}}\frac{\lambda^2}{\theta_{\rm b}^2 A_{\rm e}}\sqrt{\frac{A_{\rm S}}{\theta_{\rm b}^2}},
\end{align}
where $T_{\mathrm{sys}}=20$ {K} is the system temperature, $\delta\nu=208.9$ kHz is the MeerKAT L-band frequency resolution, $t_{\mathrm{scan}}=2$ s is the integration time per TOD sample, $A_{\rm e}$ is the effective collecting area of one dish, $A_{\mathrm S}$ is the survey area, and $\theta_{\rm b}$ is the full width at half maximum (FWHM) of the beam. The frequency dependence of $\theta_{\rm b}$ is important for the intensity-mapping simulation. Although the FWHM can be obtained directly by integrating the beam pattern at each frequency, we use the ripple model of \citep{Matshawule2021MeerKATBeamRipple}. This eighth-order polynomial model accurately describes the MeerKAT L-band FWHM:
\begin{align} 
    \theta_{b}=\frac{c}{\nu D} \left[ \sum_{d=0}^8 a_d \nu^d + A_{\rm rip}\,{\rm sin}\left( \frac{2\pi\nu}{T_{\rm rip}}\right)\right],
\end{align}
where $D=13.5$ m is the MeerKAT dish diameter, and $A_{\rm rip}=0.1$ arcmin and $T_{\rm rip}=10$ MHz are the amplitude and period of the ripple, respectively. The fitted coefficients are $a_{0}=6.7\times 10^3$, $a_{1}=-50.3$, $a_{2}=0.16$, $a_{3}=-3.0\times10^{-4}$, {$a_{4}=3.5\times10^{-7}$}, $a_{5}=-2.6\times10^{-10}$, $a_{6}=1.2\times10^{-13}$, $a_{7}=-3.0\times10^{-17}$, $a_{8}=3.4\times10^{-21}$.

After simulating the sky and noise contributions, we combine them to obtain the observed brightness temperature in each TOD sample. We adopt an angular pixel size of $0.15^\circ$ for the intensity maps \citep{Cunnington2022MeerKAT19,Meerklass2025} and reconstruct the maps from the TOD using
\begin{align}
    T_{\mathrm{obs}}
    =
    \left(\mathrm{P}^{\mathsf T}\mathrm{N}^{-1}\mathrm{P}\right)^{-1}
    \mathrm{P}^{\mathsf T}\mathrm{N}^{-1}T_{\mathrm D},
\end{align}
where $\mathrm{P}$ is the pointing matrix and $\mathrm{N}$ is the diagonal time-domain noise covariance matrix.  Figure~\ref{fig:hi_signal_noise_slice} shows the central frequency slice of the H\textsc{i} signal and thermal-noise cubes, together with the corresponding TOD number-count map. Because the effective integration time is $t_{\mathrm{tot}}=N_{\mathrm{TOD}}t_{\mathrm{scan}}$, the spatially non-uniform number of TOD samples produced by the on-the-fly scan leads directly to inhomogeneous thermal noise across the map.

\section{Projected Squared Fields for Cross-correlation}
\label{sec:hi_projected_fields}

After generating the kSZ and H\textsc{i} mock maps, we construct the squared fields used to estimate the cross-power spectrum. This preprocessing includes downgrading the angular resolution of the kSZ map and projecting the three-dimensional H\textsc{i} cube along the line of sight.

\subsection{kSZ squared fields}

Since statistical isotropy gives equal weight to objects with positive and negative line-of-sight peculiar velocities, the direct two-point correlation of the kSZ temperature with a density-like tracer vanishes in the limit of sufficient statistics. To solve this problem, squaring the kSZ field can avoid this cancellation \citep{Dore2004,Hill2016,Yuwen2025}.

To cross-correlate the kSZ and H\textsc{i} maps pixel by pixel, we downgrade the kSZ map to the angular resolution of the H\textsc{i} map using block averaging, {which means we take the average value of all the high-resolution pixels covered by the low-resolution pixel to be the value of pixel for the downgraded map. Block average can be expressed as 
\begin{align}
    T_{\rm LR}
    =
    \frac{1}{N_{\rm in}}\sum_{i\in{\rm LR}}T_{\rm{HR},i},
\end{align}
where $T_{\rm LR}$ and $T_{\rm{HR}}$ are the temperature of low-resolution map and high-resolution map, respectively. And $N_{\rm in}$ is the number of high-resolution pixels coverd by a low-resolution pixel. }

{The construction of kSZ squared field including downgrading and squaring. However, the order of these two operation matters because the resulting maps of two possible orderings are not equivalent.}
We therefore consider two constructions:
\begin{itemize}
    \item $T_{{\mathrm{kSZ}}^2}^{\mathrm{DS}}(\hat{n})$, where ``DS'' denotes downgrade then square: the kSZ map is first downgraded to the H\textsc{i} angular resolution and is then squared,
    \begin{align}
        T_{{\mathrm{kSZ}}^2}^{\mathrm{DS}}(\hat{n})
        =
        \left(\frac{1}{N_{\rm in}}\sum_{i\in\hat{n}}T_{{\rm kSZ},i}\right)^2;
    \end{align}
    
    \item $T_{{\mathrm{kSZ}}^2}^{\mathrm{SD}}(\hat{n})$, where ``SD'' denotes square then downgrade: the high-resolution kSZ map is first squared and is then downgraded to the H\textsc{i} angular resolution,
    \begin{align}
        T_{{\mathrm{kSZ}}^2}^{\mathrm{SD}}(\hat{n})
        =
        \frac{1}{N_{\rm in}}\sum_{i\in\hat{n}}T_{{\rm kSZ},i}^2.
    \end{align}
\end{itemize}
The two choices have different implications for the cross-correlation signal-to-noise ratio. In the DS construction, block averaging reduces the effective noise in each low-resolution pixel, analogous to increasing the integration time per pixel. However, averaging also causes cancellation between positive and negative kSZ fluctuations and discards small-scale kSZ information before the field is squared. In the SD construction, squaring first prevents this sign cancellation and allows small-scale fluctuations to contribute to the low-resolution squared field. The same operation, however, also retains the variance of high-resolution noise and foreground residuals and couples this small-scale contamination into larger-scale modes. We therefore construct maps and estimate power spectra for both orderings.

\begin{figure} 
    \centering
    \begin{subfigure}{0.31\textwidth}
        \centering
        \includegraphics[width=\linewidth]{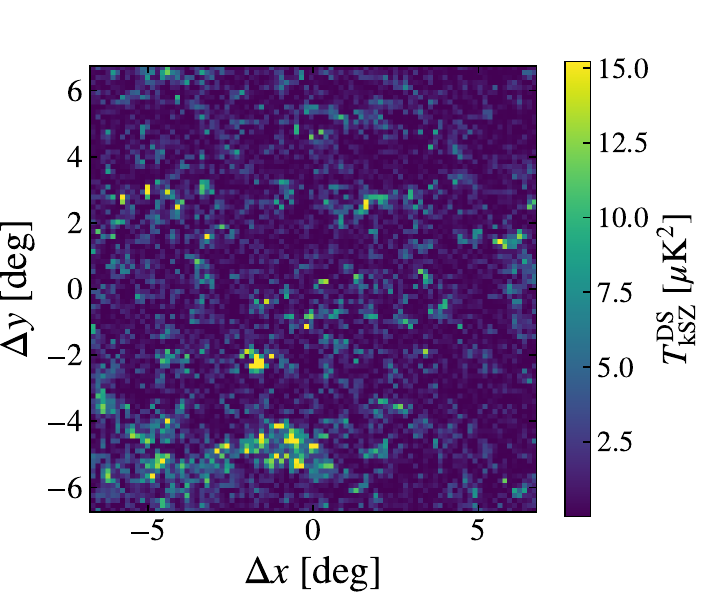}
    \end{subfigure}
    \hfill
    \begin{subfigure}{0.31\textwidth}
        \centering
        \includegraphics[width=\linewidth]{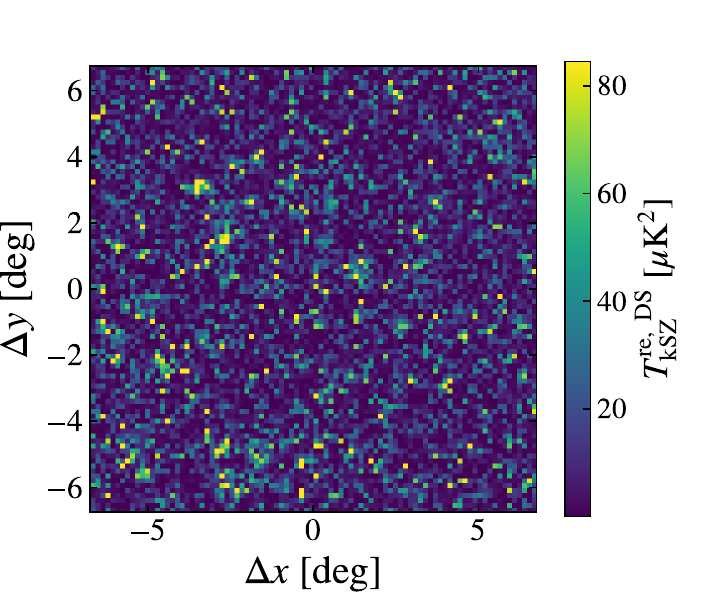}
    \end{subfigure}
    \hfill
    \begin{subfigure}{0.31\textwidth}
        \centering
        \includegraphics[width=\linewidth]{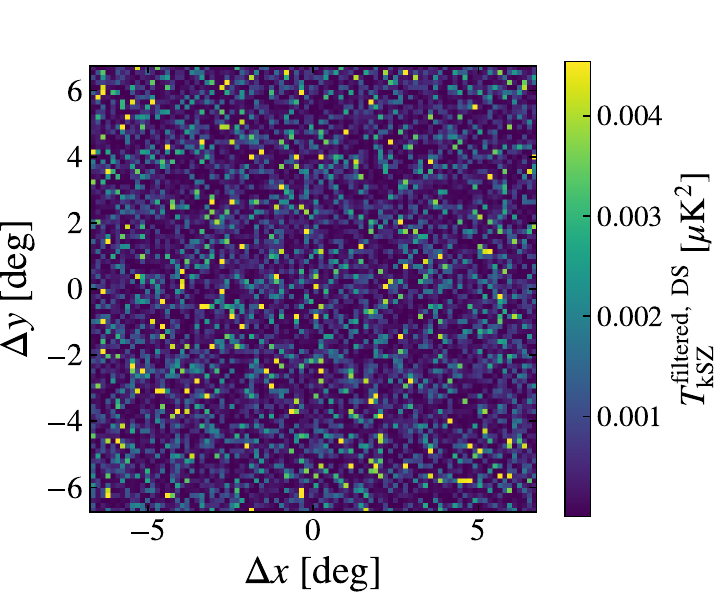}
    \end{subfigure}
    \vspace{0.5em}
    
    \begin{subfigure}{0.31\textwidth}
        \centering
        \includegraphics[width=\linewidth]{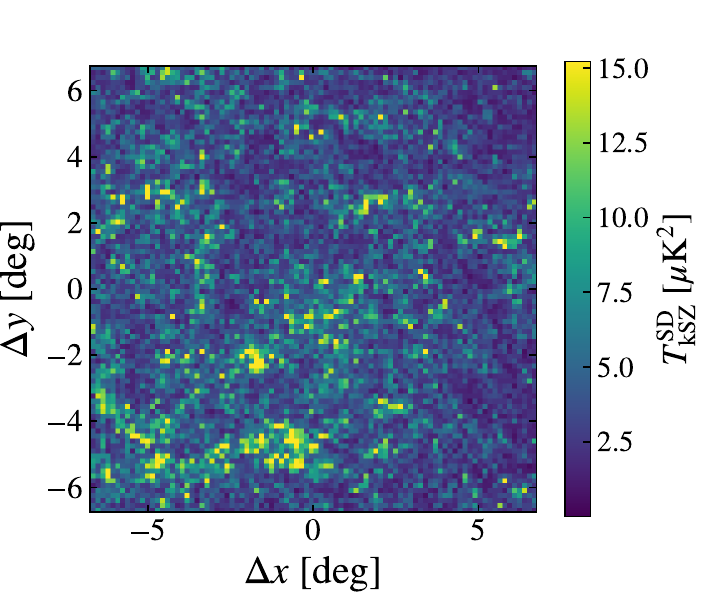}
    \end{subfigure}
    \hfill
    \begin{subfigure}{0.31\textwidth}
        \centering
        \includegraphics[width=\linewidth]{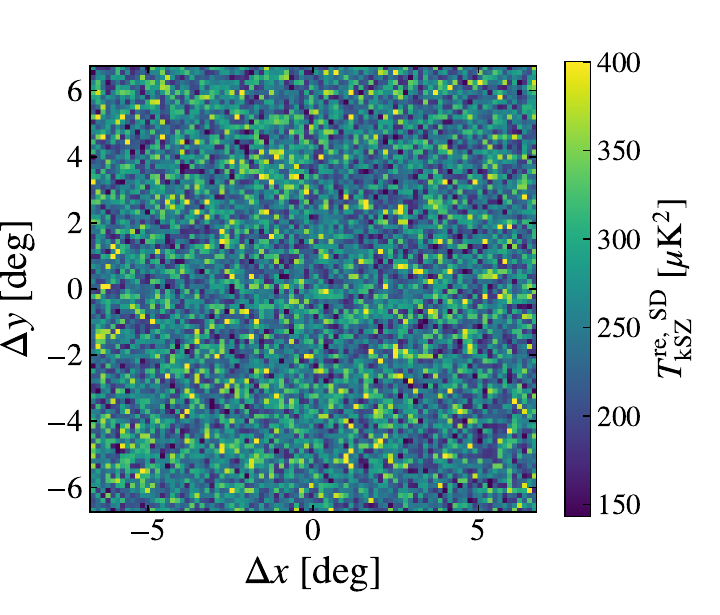}
    \end{subfigure}
    \hfill
    \begin{subfigure}{0.31\textwidth}
        \centering
        \includegraphics[width=\linewidth]{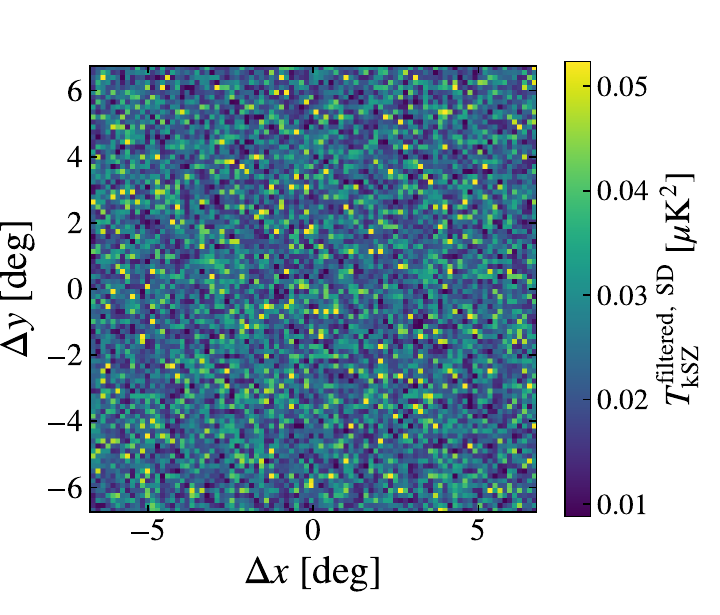}
    \end{subfigure}
    \caption{Low-resolution kSZ squared fields constructed from three high-resolution kSZ maps. The top row shows the DS construction, where the high-resolution map is first block-averaged to the H\textsc{i} angular resolution and then squared. The bottom row shows the SD construction, where the high-resolution map is squared before block averaging. From left to right, the columns show the cases of the signal-only kSZ field, the amplitude-reconstructed field $T_{\rm kSZ}^{\rm re}$, and the Wiener-filtered kSZ field $T_{\rm kSZ}^{\rm filtered}$.}
    \label{fig:ksz_squared_field_ordering_maps}
\end{figure}

Figure~\ref{fig:ksz_squared_field_ordering_maps} illustrates these constructions at map level. For the kSZ-only and Wiener-filtered fields, $T_{{\rm kSZ}^2}^{\rm SD}$ has a higher mean level and stronger small-scale contrast than $T_{{\rm kSZ}^2}^{\rm DS}$ since it retains the high-resolution variance within each block before averaging. The difference is substantially greater for the amplitude-reconstructed kSZ field, which has the highest noise level. In this case, $T_{{\rm kSZ}^2}^{\rm SD}$ retains the high-resolution CMB-noise variance as a positive contribution, whereas $T_{{\rm kSZ}^2}^{\rm DS}$ reduces part of that variance through block averaging before squaring the low-resolution map. These differences motivate treating the two orderings as distinct analysis choices rather than interchangeable numerical implementations.

\subsection{Projected squared HI fields}

The mock H\textsc{i} intensity map is a three-dimensional data cube and must be projected along the line of sight before it can be cross-correlated with the two-dimensional kSZ map. As for the kSZ field, the order of projection and squaring matters. As discussed in the previous work \citep{Yuwen2025}, projecting the squared H\textsc{i} cube retains contributions from both $k^\parallel$ and $k^\perp$ modes and therefore produces a stronger signal than squaring an already projected field. We construct the two-dimensional H\textsc{i} squared field as
\begin{equation}
    T_{\mathrm{H\textsc{i}^2}}^{\rm proj}(\hat{\bm n})
    =
    \sum_i
    W_i\,T_{\rm obs}^2(\hat{\bm n},\nu_i),
    \label{eq:hi_sq_proj_discrete}
\end{equation}
where the window function $W_i$ approximates the redshift-window normalization adopted in the analytical forecast \citep{Yuwen2025}:
\begin{equation} 
    W_i
    =
    \Delta\nu\,
    \frac{\nu_{21}}{\nu_i^2}\,
    \frac{1}{z_{\max}-z_{\min}}.
    \label{eq:hi_window}
\end{equation}
Here, $\nu_{21}=1420.4\,{\rm MHz}$ is the rest-frame 21 cm frequency, while $z_{\min}=0.40$ and $z_{\max}=0.58$ are the redshift limits of the adopted MeerKAT L-band frequency range.

\begin{figure} 
    \centering
    \begin{subfigure}{0.48\textwidth}
        \centering
        \includegraphics[width=\linewidth]{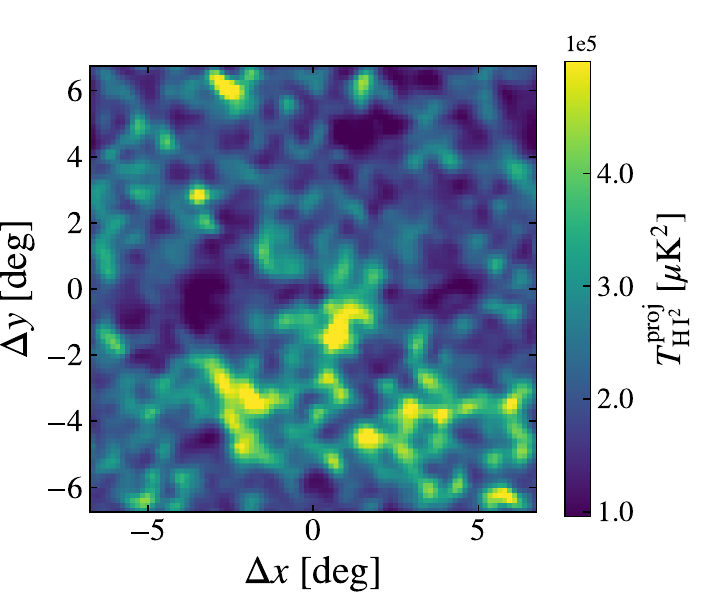}
    \end{subfigure}
    \hfill
    \begin{subfigure}{0.48\textwidth}
        \centering
        \includegraphics[width=\linewidth]{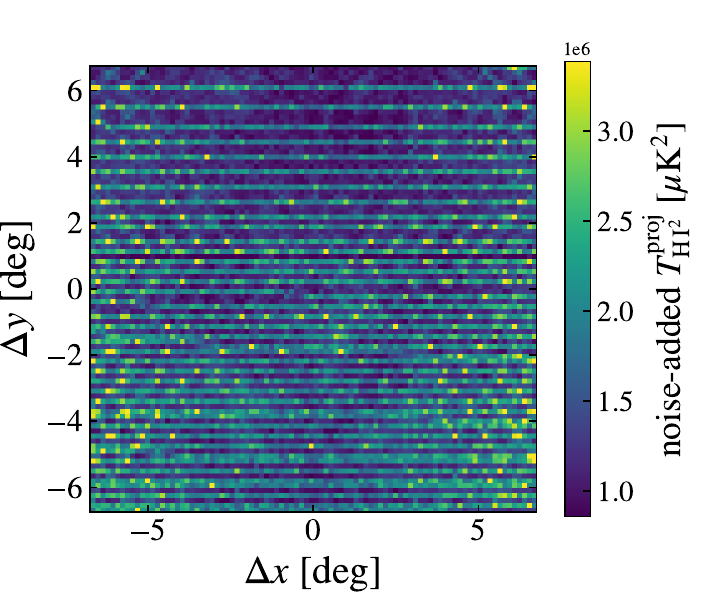}
    \end{subfigure}
    \caption{Projected squared H\textsc{i} fields, $T_{\mathrm{H\textsc{i}^2}}^{\rm proj}$, used in the kSZ--H\textsc{i} cross-correlation analysis. The left panel shows the signal-only field, and the right panel shows the field constructed from the thermal-noise-added mock MeerKAT observation.}
    \label{fig:projected_hi_squared_fields}
\end{figure}

Figure~\ref{fig:projected_hi_squared_fields} shows the projected squared H\textsc{i} fields constructed from the simulated intensity-mapping data. The left panel uses the signal-only H\textsc{i} field, and the right panel applies the same operation to the thermal-noise-added mock MeerKAT observation. The stripe-like features in the latter arise from the inhomogeneous thermal-noise variance induced by the survey scanning strategy. In each frequency channel, the thermal noise has zero mean but a position-dependent variance that scales approximately as $\sigma_{\rm H\textsc{i}}^2(\hat{\bm n},\nu)\propto1/N_{\rm TOD}(\hat{\bm n},\nu)$, where $\sigma_{\rm H\textsc{i}}^2$ is the noise variance in the voxel at $(\hat{\bm n},\nu)$. Because the thermal-noise fluctuations are substantially larger than the H\textsc{i} fluctuations in the noise-added cube, squaring gives
\begin{equation}
    \left(T_{\rm H\textsc{i}}+n_{\rm H\textsc{i}}\right)^2
    =
    T_{\rm H\textsc{i}}^2
    +2T_{\rm H\textsc{i}}n_{\rm H\textsc{i}}
    +n_{\rm H\textsc{i}}^2 ,
    \label{eq:noise_sq_proj}
\end{equation}
where $n_{\rm H\textsc{i}}$ is the MeerKAT thermal-noise brightness temperature in each voxel. The final term dominates the noise-added squared map and has a non-zero expectation value, $\langle n_{\rm H\textsc{i}}^2\rangle=\sigma_{\rm H\textsc{i}}^2$. Consequently, the projected noise-added H\textsc{i} squared field traces the spatially varying TOD number-count pattern.

\section{Power-spectrum Estimation and SNR Prediction}
\label{sec:spectra}

\subsection{Power spectrum estimation}

After constructing the squared fields, we estimate their angular auto- and cross-power spectra. Because the simulated survey covers a relatively small sky patch ($<200\,\mathrm{deg}^2$), we adopt the flat-sky approximation. In this limit, spherical-harmonic basis functions can be approximated locally by two-dimensional plane waves,
\begin{equation}
    Y_{\ell m} \sim e^{i\bm{k} \cdot \bm{\theta}} (\ell \simeq k),
\end{equation}
and the Fourier-space representation of a field is
\begin{equation}
    \tilde F(\boldsymbol{\ell})
    =\sum_{\bm{\theta}} F(\boldsymbol{\theta}) e^{-i\boldsymbol{\ell}\cdot\boldsymbol{\theta}}.
\end{equation}
The Fourier transforms of the kSZ and H\textsc{i} squared fields can then be written as 
\begin{equation}
    \tilde T_{{\rm kSZ}^2}^{\rm DS}(\bm\ell)
    = 
    \left[W_{\ell}^{\rm ba}B_\ell^{\rm A} \tilde \delta_{\rm kSZ}(\bm\ell)\right] 
    \ast
    \left[W_{\ell}^{\rm ba}B_\ell^{\rm A} \tilde \delta_{\rm kSZ}(\bm\ell)\right],
    \label{eq:FT_kSZ_DS}
\end{equation}
\begin{equation}
    \tilde T_{{\rm kSZ}^2}^{\rm SD}(\bm\ell)
    = W_{\ell}^{\rm ba}
    \left\{\left[B_\ell^{\rm A} \tilde \delta_{\rm kSZ}(\bm\ell)\right] 
    \ast
    \left[B_\ell^{\rm A} \tilde \delta_{\rm kSZ}(\bm\ell)\right] \right\},
    \label{eq:FT_kSZ_SD}
\end{equation}
\begin{equation}
    \tilde T_{{\rm H\textsc{i}}^2}^{\rm proj}(\bm\ell)
    = 
    \left[W_{\ell}^{\rm pix}B_\ell^{\rm M} \tilde \delta_{\rm H\textsc{i}}(\bm\ell)\right] 
    \ast
    \left[W_{\ell}^{\rm pix}B_\ell^{\rm M} \tilde \delta_{\rm H\textsc{i}}(\bm\ell)\right],
    \label{eq:FT_HI}
\end{equation}
where $B_\ell^{\rm A}$ and $B_\ell^{\rm M}$ are the ACT and MeerKAT beam responses, respectively, $W_{\ell}^{\rm ba}$ is the block-averaging response, $W_{\ell}^{\rm pix}$ is the intensity-map pixel response, and ``$\ast$'' denotes convolution. 
We estimate the auto- and cross-power spectra as
\begin{equation}
    C_{\ell}^{\mathrm{kSZ}^2}
    = \frac{\Omega_{\rm survey}}{N_{\rm pix}^2}
    \left\langle
    {\rm Re}
    \left[
    \tilde T_{\mathrm{kSZ}^2}(\boldsymbol{\ell})
    \tilde T_{\mathrm{kSZ}^2}^*(\boldsymbol{\ell})
    \right]
    \right\rangle ,
\end{equation}
\begin{equation}
    C_{\ell}^{\mathrm{H\textsc{i}}^2}
    = \frac{\Omega_{\rm survey}}{N_{\rm pix}^2}
    \left\langle
    {\rm Re}
    \left[
    \tilde T_{\mathrm{H\textsc{i}}^2}(\boldsymbol{\ell})
    \tilde T_{\mathrm{H\textsc{i}}^2}^*(\boldsymbol{\ell})
    \right]
    \right\rangle ,
\end{equation}
\begin{equation}
    C_{\ell}^{\mathrm{kSZ}^2 \times \mathrm{H\textsc{i}}^2}
    = \frac{\Omega_{\rm survey}}{N_{\rm pix}^2}
    \left\langle
    {\rm Re}
    \left[
    \tilde T_{\mathrm{kSZ}^2}(\boldsymbol{\ell})
    \tilde T_{\mathrm{H\textsc{i}}^2}^*(\boldsymbol{\ell})
    \right]
    \right\rangle ,
\end{equation}
where $\Omega_{\rm survey}$ is the survey area and $N_{\rm pix}$ is the total number of pixels. The power-spectrum uncertainties are estimated as
\begin{equation}
    \sigma_{\ell}^{\mathrm{kSZ}^2}
    = \frac{1}{\sqrt{N_{\rm m}(\ell)}}
    \left|C_{\ell}^{\mathrm{kSZ}^2}\right|,
\end{equation}
\begin{equation}
    \sigma_{\ell}^{\mathrm{H\textsc{i}}^2}
    = \frac{1}{\sqrt{N_{\rm m}(\ell)}}
    \left|{C_{\ell}^{\mathrm{H\textsc{i}}^2}}\right|,
\end{equation}
\begin{equation}
   {\sigma_{\ell}^{\mathrm{kSZ}^2 \times \mathrm{H\textsc{i}}^2}
    = \frac{1}{\sqrt{2N_{\rm m}(\ell)}}
    \sqrt{\left(C_{\ell}^{\mathrm{kSZ}^2 \times \mathrm{H\textsc{i}}^2}\right)^2+C_{\ell}^{\mathrm{H\textsc{i}}^2}C_{\ell}^{\mathrm{kSZ}^2}},}
\end{equation}
where the angle brackets denote averages over the Fourier modes in each annular bin and $N_{\rm m}(\ell)$ is the corresponding number of independent modes.



\begin{figure*}[!htbp]
    \centering
    \begin{tabular}{cc}
        \includegraphics[width=0.47\textwidth]{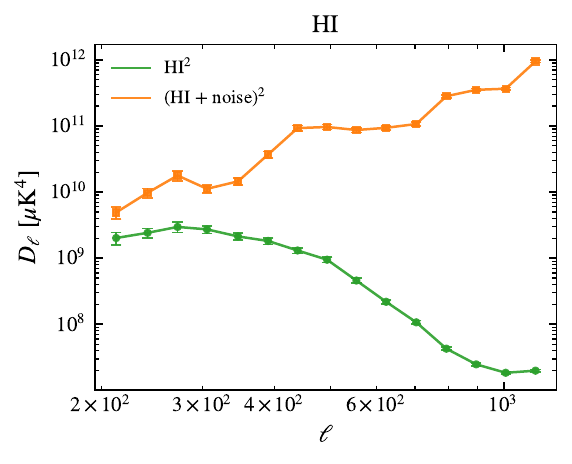} &
        \includegraphics[width=0.47\textwidth]{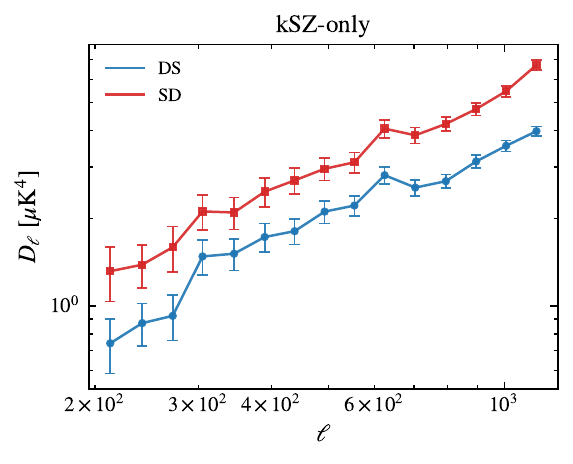} \\
        \includegraphics[width=0.47\textwidth]{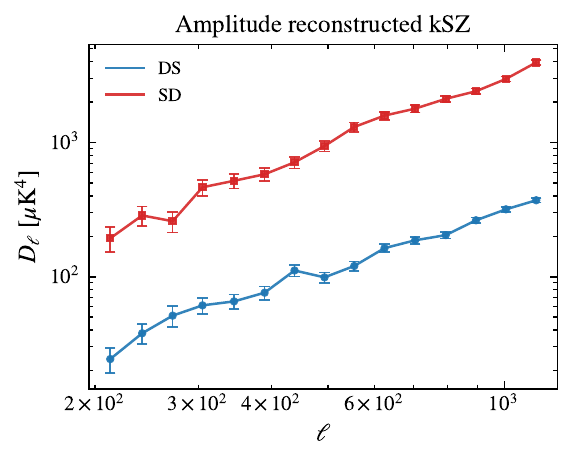} &
        \includegraphics[width=0.47\textwidth]{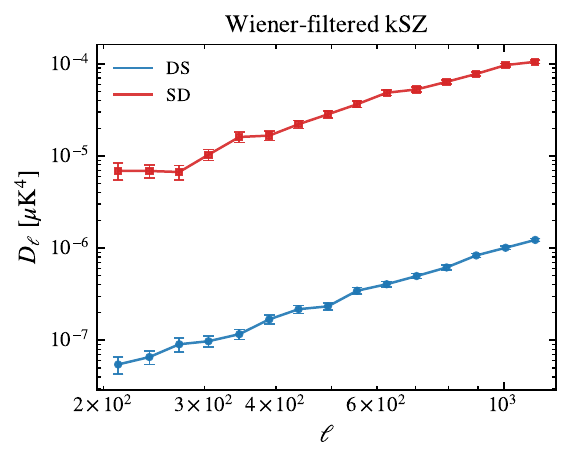}
    \end{tabular}
    \caption{Auto-spectra of the squared fields used in the cross-correlation analysis. The upper-left panel compares the projected squared H\textsc{i} fields constructed from the signal-only and noise-added maps. The upper-right, lower-left, and lower-right panels show the squared-field auto-spectra constructed from the beam-smoothed kSZ, amplitude-reconstructed kSZ, and Wiener-filtered kSZ maps, respectively. In each kSZ panel, the DS and SD curves denote the downgrade-then-square and square-then-downgrade constructions.}
    \label{fig:squared_field_auto_spectra}
\end{figure*}

Before presenting the cross-correlation measurements, we inspect the auto-spectra of the squared fields entering the estimator. Figure~\ref{fig:squared_field_auto_spectra} shows the auto-spectrum of the projected squared H\textsc{i} field and those of the squared fields constructed from the three kSZ map variants considered in this work. These comparisons illustrate how instrumental noise and Wiener filtering modify the variance of the squared kSZ maps before cross-correlation with the projected H\textsc{i} squared field.
The noise-added projected H\textsc{i} squared field has substantially larger power than its signal-only counterpart because squaring converts the thermal-noise variance into a non-zero mean field. 
As for the auto-spectra of kSZ square fields, the DS and SD constructions lead to similar spectral shapes for all three kSZ-based squared fields, but SD construction always have higher overall amplitudes. This is because The SD construction maintains variance of small scale by squaring before averaging.
The amplitude difference between SD and DS becomes larger when ILC noise is added to the kSZ map, since the variance of thermal noise also contribute to the intra-pixel variance retained by the SD construction.
For the Wiener-filtered case, the difference between the DS and SD auto-spectra is amplified, which is somewhat counter-intuitive because the Wiener filter is supposed to downweight the contamination from primary CMB and ILC noise. A possible explanation is that the filter strongly reshapes the scale dependence of the input map. The subsequent block average suppresses the remaining intra-pixel fluctuations in the DS construction, whereas the SD construction converts these fluctuations into a positive local variance before downgrading. The resulting auto-power therefore depends sensitively on the ordering of filtering, squaring, and angular downgrading.

\begin{figure*}[!htbp] 
    \centering
    \includegraphics[width=0.32\textwidth]{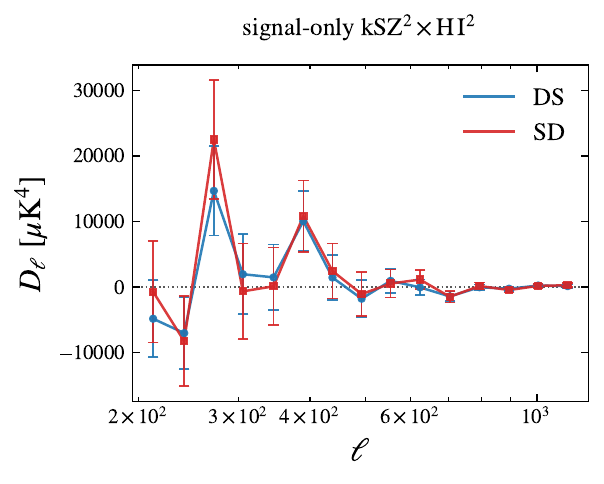}
    \includegraphics[width=0.32\textwidth]{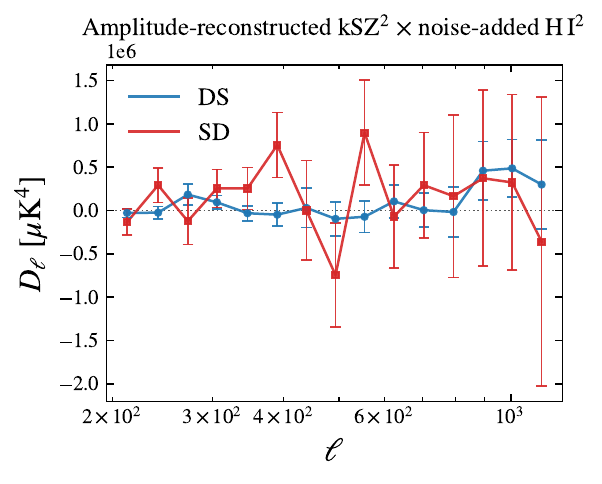}
    \includegraphics[width=0.32\textwidth]{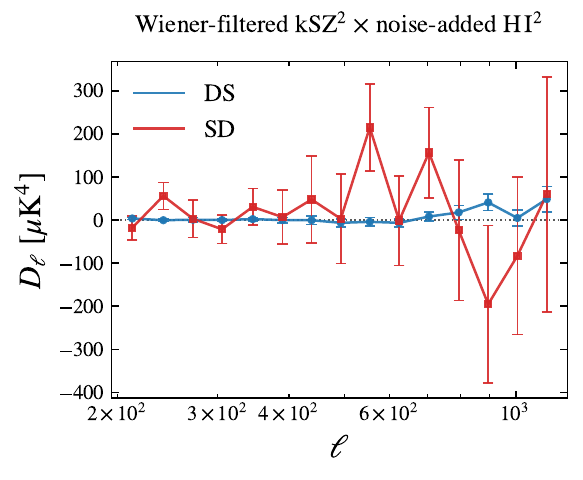}
    \caption{Cross-power spectra between the projected H\textsc{i} squared field and the kSZ squared fields. 
    The left panel uses the signal-only $\rm kSZ^2$ and $\rm H\textsc{i}^2$ maps.  
    The middle and right panels cross-correlate the noise-added $\rm H\textsc{i}^2$ map with the amplitude-reconstructed $\rm kSZ^2$ and Wiener-filtered $\rm kSZ^2$ maps, respectively. 
    In each panel, the DS and SD curves correspond to the downgrade-then-square and square-then-downgrade constructions.}
    \label{fig:squared_field_cross_spectra}
\end{figure*}

We next compute the cross-power spectra between the projected H\textsc{i} squared field and the corresponding kSZ squared fields. In Figure~\ref{fig:squared_field_cross_spectra}, we compare three cases: the signal-only $\rm kSZ^2$--$\rm H\textsc{i}^2$ cross-spectrum, the cross-spectrum of the amplitude-reconstructed $\rm kSZ^2$ field with the noise-added $\rm H\textsc{i}^2$ field, and the cross-spectrum of the Wiener-filtered $\rm kSZ^2$ field with the noise-added $\rm H\textsc{i}^2$ field.
In the signal-only case, the DS and SD constructions have broadly similar scale dependence, although the SD spectrum shows a slightly larger amplitude. This suggests that, in the absence of observational contaminants, the unresolved kSZ variance within a low-resolution H\textsc{i} pixel does not qualitatively change the cross-correlation spectrum. 
The cross-spectra of amplitude-reconstructed $\rm kSZ^2$ $\times$ noise-added $\rm H\textsc{i}^2$ shows that, the scale-dependency of amplitude differs from the signal-only cross-spectra because of the entering of noise. This behavior indicates that the noise contribution in the square field cannot be simply removed by the cross-correlation. And the DS and SD spectra show larger differences in both amplitude and scale dependence, indicating that the behavior of noise in the cross-correlation power spectrum is strongly affected by different square field construction. 
For the Wiener-filtered $\rm kSZ^2$ $\times$ noise-added $\rm H\textsc{i}^2$, the overall amplitude is reduced relative to the noise-added case, consistent with the filter suppressing modes dominated by the primary CMB and ILC noise. Nevertheless, the DS and SD spectra remain visibly different, and their scale dependence can differ even more strongly than in the unfiltered noisy case. This may arise for the same reason discussed for the squared-field auto-spectra: Wiener filtering reshapes the scale dependence of the input map, so the remaining fluctuations can be more sensitive to whether squaring is applied before or after angular downgrading.

\subsection{SNR prediction}
To assess the detectability, we further estimate the signal-to-noise ratio (SNR) of the kSZ--H\textsc{i} cross-correlation. For the diagonal error model adopted here, the squared cumulative SNR is defined as

\begin{equation}
    \left(\frac{S}{N}\right)^2
    =
    \sum_{\ell}
    \left(\frac{C_{\ell}}{\sigma_{\ell}}\right)^2,
    \label{eq:chi2}
\end{equation}

\begin{table}
\begin{center}
\caption{SNRs of the squared-field cross-power spectra.}
\label{tab:snr}
\begin{tabular}{lcc}
\hline\noalign{\smallskip}
Case & DS kSZ & SD kSZ \\
\hline\noalign{\smallskip}
Signal-only $\rm kSZ^2$ $\times$ $\rm H\textsc{i}^2$ & 4.3 & 4.2 \\
Amplitude-reconstructed $\rm kSZ^2$ $\times$noise-added $\rm H\textsc{i}^2$ & 3.1 & 3.7 \\
Wiener-filtered $\rm kSZ^2$ $\times$noise-added $\rm H\textsc{i}^2$ & 3.6 & 3.6 \\
\noalign{\smallskip}\hline
\end{tabular}
\end{center}
\end{table}

Table~\ref{tab:snr} summarizes the cumulative SNRs obtained from the squared-field cross-spectra. In the signal-only case, the two ordering choices give very similar values: ${\rm SNR}=4.3$ for $T_{{\rm H\textsc{i}}^2}^{\rm proj}\times T_{{\mathrm{kSZ}}^2}^{\mathrm{DS}}$ and ${\rm SNR}=4.2$ for $T_{{\rm H\textsc{i}}^2}^{\rm proj}\times T_{{\mathrm{kSZ}}^2}^{\mathrm{SD}}$. 
The SD construction retains additional within-pixel kSZ variance and therefore produces a slightly different cross-power spectrum, but this difference does not lead to a comparable change in the cumulative SNR. As shown by the analytic estimate in Appendix~\ref{app:ds_sd_snr}, the two squared-field constructions retain different scale-dependent cross-correlation information while providing similar cumulative detectability.

After adding ACT ILC noise and MeerKAT thermal noise to the mock signals, the difference between the two squared-field orderings increases. In this case, it gives ${\rm SNR}=3.1$ for $T_{{\rm H\textsc{i}}^2}^{\rm proj}\times T_{{\mathrm{kSZ}}^2}^{\mathrm{DS}}$ and ${\rm SNR}=3.7$ for $T_{{\rm H\textsc{i}}^2}^{\rm proj}\times T_{{\mathrm{kSZ}}^2}^{\mathrm{SD}}$. The overall reduction in SNR relative to the signal-only case is qualitatively expected. 
However, the square-then-downgrade construction has higher SNR is unexpected. 
Qualitatively speaking, we assume the reason of this behavior may be a contribution from false correlation caused by the high-resolution kSZ ILC noise entering the cross-correlation power spectrum with H\textsc{i}.
If future component-separation methods recover the kSZ field independently from the total CMB temperature fluctuations, residual noise and foregrounds will therefore require particular care when the high-resolution map is squared before downgrading. 

For the Wiener-filtered case, the results are ${\rm SNR}=3.6$ for $T_{{\rm H\textsc{i}}^2}^{\rm proj}\times T_{{\mathrm{kSZ}}^2}^{\mathrm{DS}}$ and ${\rm SNR}=3.6$ for $T_{{\rm H\textsc{i}}^2}^{\rm proj}\times T_{{\mathrm{kSZ}}^2}^{\mathrm{SD}}$. The SNR of two orderings are almost indistinguishable while their cross-correlation power spectra have visible difference. This situation could also be explained by the analytical derivation in Appendix~\ref{app:ds_sd_snr}. 
Taken together, the signal-only and Wiener-filtered results, along with the analytic estimate in Appendix~\ref{app:ds_sd_snr}, show that similar cumulative SNRs do not imply that the DS and SD cross-spectra are statistically redundant. The different ordering of squaring and angular downgrading preserves different scale-dependent aspects of the cross-correlation, suggesting that the two constructions may provide complementary information. A joint analysis of both estimators, with their cross-covariance properly taken into account, may therefore recover more of the available cross-correlation information than either construction alone. We leave a detailed investigation of this possibility to future work. 

Overall, the squared-field estimator yields cumulative $\mathrm{kSZ}^2$--$\mathrm{H\textsc{i}}^2$ cross-correlation SNRs of approximately $3$--$4$, demonstrating the potential detectability of this statistic in future joint ACT and MeerKAT observations. 


\section{Summary and Discussion}
\label{sec:summary_discussion}

We have presented a map-level simulation study of the cross-correlation between the squared kSZ field and the projected squared H\textsc{i} intensity field. Both observables are constructed from the same Jiutian cosmological lightcone, thereby preserving their shared underlying density and velocity fields. 
The kSZ maps are generated from a halo-based electron distribution model and are combined with lensed primary CMB and effective ILC noise to construct amplitude-reconstructed and Wiener-filtered kSZ maps that approximate idealized ACT observations. The H\textsc{i} data cube is generated with H\textsc{i} mass in lightcone and 
a MeerKAT-like survey setup over the same sky patch.

We construct the H\textsc{i} squared field by squaring the three-dimensional data cube and then projecting it along the line of sight. Because the CMB map is sampled much more finely than the H\textsc{i} map, we compare two constructions of the low-resolution kSZ squared field. In the DS case, block averaging suppresses unresolved signal and noise before the nonlinear operation. In the SD case, the high-resolution variance is retained before averaging. We then estimate the auto- and cross-power spectra of the resulting squared fields.

We calculate the SNR of the cross-spectra to assess the detectability of the $\mathrm{kSZ}^2\times\mathrm{H\textsc{i}}^2$ correlation. 
The signal-only cross-correlation gives similar values, ${\rm SNR}=4.3$ for DS and $4.2$ for SD, which indicates unresolved kSZ variance within a $0.15^\circ$ H\textsc{i} pixel not contribute much to the cumulative statistic. 
For the amplitude-reconstructed case, the diagnostic values are ${\rm SNR}=3.1$ for DS and $3.7$ for SD. The larger SNR in the SD branch may indicate that the high-resolution kSZ ILC noise in the square field may cause a false signal in the $\rm kSZ^2$-$\rm {H\textsc{i}}^2$ correlation. 
By contrast, the Wiener-filtered branch gives nearly identical values, ${\rm SNR}=3.6$ and $3.6$ for DS and SD, respectively. 
In both cases of signal-only and Wiener-filtered branch, the DS and SD constructions retain different cross-correlation power across angular scales but yield similar cumulative detectability. This behavior can be understood through our derivation shown in Appendix~\ref{app:ds_sd_snr}, and suggests that cross-correlation of different construction may have similar SNR while they contains different statistical information. 

Overall, the squared-field estimator yields cumulative SNR of $\rm kSZ^2$-$\rm H\textsc{i}^2$ cross-correlation at the level of $S/N\simeq3$--$4$, which validates the detectability of future joint observation of ACT and MeerKAT. 
Owing to the idealized assumptions of the simulation, these results should not be interpreted as the significance attainable with real ACT and MeerKAT data. Our simulation didn't include the foreground residual of kSZ map and the foreground removal and signal compensation in the square field of H\textsc{i} intensity mapping, whose effects on the SNR will be further studied in our futhre work. 
Nevertheless, the overlap survey area of future ACT and MeerKAT or future SKA H\textsc{i} intensity-mapping surveys could exceed that of our simulated field by more than an order of magnitude, the actual observed SNR can still reach a detectable level. 
Our results indicate that the kSZ--H\textsc{i} squared-field correlation is a promising observable for future joint CMB and 21 cm intensity-mapping analyses. 
A successful low-redshift measurement would provide a new cross-check of the baryonic density and velocity fields and help establish the methodology required to apply related 21 cm--kSZ statistics over broader redshift ranges.

\acknowledgments
This work makes use of Astropy \citep{2013A&A...558A..33A,2018AJ....156..123A,2022ApJ...935..167A}, NumPy, SciPy, and Matplotlib.
This work is supported by the CAS Project for Young Scientists in Basic Research (No. YSBR-92), National Key R\&D Program of China grant Nos. 2022YFF0503404, the science research grants from the China Manned Space Project with grant Nos. CMS-CSST-2025-A02. Y.E.J. and Z.Y.Y are supported by the Program of China Scholarship Council Grant No. 202404910398 and No. 202404910329. Y.-Z. Ma acknowledges South Africa Research Chair Initiative (SARChI; Grant No. RCCA250325306198) from South Africa's Department of Science, Technology and Innovation and National Research Foundation (NRF), and other NRF Grants with No.~150580, No.~CHN22111069370, No.~ERC250324306141, No. AUPP250310302114. The Jiutian simulations were conducted under the support of the science research grants from the China Manned Space Project with grant No. CMS- CSST- 2021-A03.




\appendix

\section{Analytic comparison of the DS and SD signal-to-noise ratios}
\label{app:ds_sd_snr}

To clarify why the downgrade-then-square (DS) and square-then-downgrade
(SD) constructions can yield similar cumulative signal-to-noise ratios
(SNRs) despite their different power-spectrum amplitudes, we provide a
simple analytic estimate below.

Suppose that each low-resolution pixel centred at $\hat{\bm n}$ contains
$N_{\rm in}$ high-resolution kSZ pixels. The DS and SD squared fields are
defined as
\begin{align}
    T_{{\rm kSZ}^2}^{\rm DS}(\hat{\bm n})
    &=
    \left(
    \frac{1}{N_{\rm in}}
    \sum_{i\in\hat{\bm n}}T_{{\rm kSZ},i}
    \right)^2,
    \label{eq:app_ds_field}
    \\
    T_{{\rm kSZ}^2}^{\rm SD}(\hat{\bm n})
    &=
    \frac{1}{N_{\rm in}}
    \sum_{i\in\hat{\bm n}}T_{{\rm kSZ},i}^2.
    \label{eq:app_sd_field}
\end{align}
Their difference is
\begin{align}
    \Delta(\hat{\bm n})
    &\equiv
    T_{{\rm kSZ}^2}^{\rm SD}(\hat{\bm n})
    -T_{{\rm kSZ}^2}^{\rm DS}(\hat{\bm n})
    \\
    &=
    \overline{T^2}_{{\rm kSZ},\hat{\bm n}}
    -\overline{T}_{{\rm kSZ},\hat{\bm n}}^{,2},
    \label{eq:app_delta_field}
\end{align}
which is the unresolved temperature variance within the corresponding
low-resolution pixel. Consequently, $\Delta(\hat{\bm n})\geq0$ pixel by
pixel, and
\begin{equation}
    T_{{\rm kSZ}^2}^{\rm SD}(\hat{\bm n})
    =
    T_{{\rm kSZ}^2}^{\rm DS}(\hat{\bm n})
    +\Delta(\hat{\bm n}).
    \label{eq:app_field_relation}
\end{equation}

The linearity of the Fourier transform gives
\begin{equation}
    \widetilde T_{{\rm kSZ}^2}^{\rm SD}(\bm\ell)
    =
    \widetilde T_{{\rm kSZ}^2}^{\rm DS}(\bm\ell)
    +\widetilde\Delta(\bm\ell).
    \label{eq:app_fourier_relation}
\end{equation}
The corresponding SD auto-power spectrum is therefore
\begin{equation}
    C_\ell^{\mathrm{kSZ}^2,\mathrm{SD}}
    =
    C_\ell^{\mathrm{kSZ}^2,\mathrm{DS}}
    +2C_\ell^{(\mathrm{kSZ}^2,\mathrm{DS})\times\Delta}
    +C_\ell^\Delta.
    \label{eq:app_auto_exact}
\end{equation}
The relation between $C_\ell^\Delta$ and $C_\ell^{\mathrm{kSZ}^2,\mathrm{DS}}$ can be defined by a scale-dependent ratio $\alpha_\ell$ through
\begin{equation}
    C_\ell^\Delta
    =
    \alpha_\ell C_\ell^{\mathrm{kSZ}^2,\mathrm{DS}}.
    \label{eq:app_alpha_definition}
\end{equation}
Using the corresponding approximate scaling
\begin{equation}
    C_\ell^{(\mathrm{kSZ}^2,\mathrm{DS})\times\Delta}
    \simeq
    \sqrt{\alpha_\ell}
    C_\ell^{\mathrm{kSZ}^2,\mathrm{DS}},
    \label{eq:app_delta_cross_scaling}
\end{equation}
Equation~\eqref{eq:app_auto_exact} becomes
\begin{equation}
    C_\ell^{\mathrm{kSZ}^2,\mathrm{SD}}
    \simeq
    (1+\sqrt{\alpha_\ell})^2
    C_\ell^{\mathrm{kSZ}^2,\mathrm{DS}}.
    \label{eq:app_auto_scaling}
\end{equation}

For the cross-power spectrum with the projected H\textsc{i} squared field,
Equation~\eqref{eq:app_field_relation} gives
\begin{equation}
    C_\ell^{(\mathrm{kSZ}^2,\mathrm{SD})\times\mathrm{H\textsc{i}}^2}
    =
    C_\ell^{(\mathrm{kSZ}^2,\mathrm{DS})\times\mathrm{H\textsc{i}}^2}
    +C_\ell^{\Delta\times\mathrm{H\textsc{i}}^2}.
    \label{eq:app_cross_exact}
\end{equation}
With the analogous approximate scaling of the additional term, this relation
can be written as
\begin{equation}
    C_\ell^{(\mathrm{kSZ}^2,\mathrm{SD})\times\mathrm{H\textsc{i}}^2}
    \simeq
    (1+\sqrt{\alpha_\ell})
    C_\ell^{(\mathrm{kSZ}^2,\mathrm{DS})\times\mathrm{H\textsc{i}}^2}.
    \label{eq:app_cross_scaling}
\end{equation}

The variance of
the cross-power spectrum in one multipole bin is
\begin{equation}
    \left(
    \sigma_\ell^{\mathrm{kSZ}^2\times\mathrm{H\textsc{i}}^2}
    \right)^2
    =
    \frac{
    \left(
    C_\ell^{\mathrm{kSZ}^2\times\mathrm{H\textsc{i}}^2}
    \right)^2
    +C_\ell^{\mathrm{H\textsc{i}}^2}C_\ell^{\mathrm{kSZ}^2}
    }{2N_{\rm m}(\ell)},
    \label{eq:app_cross_variance}
\end{equation}
where $N_{\rm m}(\ell)$ is the number of modes used in the corresponding
multipole bin. The differential SNR of the SD construction is consequently
\begin{align}
    \left(\frac{S}{N}\right)^2_{\ell,\mathrm{SD}}
    ={}&
    2N_{\rm m}(\ell)
    \frac{
    (1+\sqrt{\alpha_\ell})^2
    \left(
    C_\ell^{(\mathrm{kSZ}^2,\mathrm{DS})\times\mathrm{H\textsc{i}}^2}
    \right)^2
    }{
    (1+\sqrt{\alpha_\ell})^2
    \left(
    C_\ell^{(\mathrm{kSZ}^2,\mathrm{DS})\times\mathrm{H\textsc{i}}^2}
    \right)^2
    +(1+\sqrt{\alpha_\ell})^2
    C_\ell^{\mathrm{H\textsc{i}}^2}
    C_\ell^{\mathrm{kSZ}^2,\mathrm{DS}}
    }
    \notag\\
    ={}&
    2N_{\rm m}(\ell)
    \frac{
    \left(
    C_\ell^{(\mathrm{kSZ}^2,\mathrm{DS})\times\mathrm{H\textsc{i}}^2}
    \right)^2
    }{
    \left(
    C_\ell^{(\mathrm{kSZ}^2,\mathrm{DS})\times\mathrm{H\textsc{i}}^2}
    \right)^2
    +C_\ell^{\mathrm{H\textsc{i}}^2}
    C_\ell^{\mathrm{kSZ}^2,\mathrm{DS}}
    }
    \notag\\
    ={}&
    \left(\frac{S}{N}\right)^2_{\ell,\mathrm{DS}}.
    \label{eq:app_snr_per_bin}
\end{align}
The common scale-dependent factor therefore cancels between the
cross-spectrum and its Gaussian uncertainty. Summing the differential SNR
over the same multipole bins gives
\begin{equation}
    \left(\frac{S}{N}\right)^2_{\mathrm{SD}}
    =
    \sum_{\ell\,\mathrm{bins}}
    \left(\frac{S}{N}\right)^2_{\ell,\mathrm{SD}}
    \simeq
    \sum_{\ell\,\mathrm{bins}}
    \left(\frac{S}{N}\right)^2_{\ell,\mathrm{DS}}
    =
    \left(\frac{S}{N}\right)^2_{\mathrm{DS}}.
    \label{eq:app_snr_cumulative}
\end{equation}
This provides a simple interpretation of the situation that two constructions of kSZ squared fields can yield similar cumulative SNR when their cross-power spectra 
retain different scale-dependent cross-correlation information.




\bibliographystyle{JHEP}
\bibliography{bibtex}

\end{document}